\documentclass[aps,pre,onecolumn,showkey,square,numbers,amssymb,amsmath,nobibnotes,nofootinbib,reprint]{revtex4}
\pdfoutput=1

\usepackage{bm}
\usepackage{dsfont}
\usepackage[utf8]{inputenc}
\usepackage{amsfonts}
\usepackage{graphicx} 
\usepackage{amsthm}
\usepackage{placeins}
\usepackage{verbatim}
\usepackage{booktabs}
\usepackage{color}
\usepackage{bm}
\usepackage{bbold}
\usepackage[colorlinks=true,linkcolor=red,citecolor=blue]{hyperref}
\usepackage{comment}

\newcommand{\be}{\begin{equation}}
\newcommand{\ee}{\end{equation}}
\newcommand{\de}{\mathrm{d}}

\begin{document}

\title[Invariant sums of random matrices and the onset of level repulsion]{Invariant sums of random matrices and the onset of level repulsion}

\author{Zdzis\l aw Burda$^{1}$, Giacomo Livan$^2$, Pierpaolo Vivo$^{3,4}$}

\affiliation{$^1$AGH University of Science and Technology,
Faculty of Physics and Applied Computer Science,
al. Mickiewicza 30, PL-30059 Krakow, Poland\\
$^2$ Department of Computer Science, University College London, Gower Street, WC1E 6BT London, United Kingdom\\
$^3$ King's College London, Department of Mathematics, Strand, London WC2R 2LS, United Kingdom\\
$^4$ On leave from Laboratoire de Physique Th\'eorique et Mod\`eles Statistiques, UMR CNRS 8626, Universit\'e Paris-Sud, 91405 Orsay, France}
\email{zdzislaw.burda@gmail.com, g.livan@ucl.ac.uk, pierpaolo.vivo@kcl.ac.uk}
\vspace{10pt}

\begin{abstract}
We compute analytically the joint probability density of eigenvalues and the level spacing statistics for an ensemble of random matrices with interesting features. It is invariant under the standard symmetry groups (orthogonal and unitary) and yet the interaction between eigenvalues is not Vandermondian. The ensemble contains real symmetric or complex hermitian matrices $\mathbf{S}$ of the form $\mathbf{S}=\sum_{i=1}^M \langle \mathbf{O}_i \mathbf{D}_i\mathbf{O}_i^{\mathrm{T}}\rangle$ or $\mathbf{S}=\sum_{i=1}^M \langle \mathbf{U}_i \mathbf{D}_i\mathbf{U}_i^\dagger\rangle$ respectively. The diagonal matrices $\mathbf{D}_i=\mathrm{diag}\{\lambda_1^{(i)},\ldots,\lambda_N^{(i)}\}$ are constructed from real eigenvalues drawn \emph{independently} from distributions $p^{(i)}(x)$, while the matrices $\mathbf{O}_i$ and $\mathbf{U}_i$ are all orthogonal or unitary. The average $\langle\cdot\rangle$ is simultaneously performed over the symmetry group and the joint distribution of $\{\lambda_j^{(i)}\}$. We focus on the limits i.) $N\to\infty$ and ii.) $M\to\infty$, with $N=2$. In the limit i.), the resulting sum $\mathbf{S}$ develops level repulsion even though the original matrices do not feature it, and classical RMT universality is restored asymptotically. In the limit ii.) the spacing distribution attains scaling forms that are computed exactly: for the orthogonal case, we recover the $\beta=1$ Wigner's surmise, while for the unitary case an entirely new universal distribution is obtained. Our results allow to probe analytically the microscopic statistics of the sum of random matrices that become asymptotically free. We also give an interpretation of this model in terms of radial random walks in a matrix space. The analytical results are corroborated by numerical simulations.
\end{abstract}

\maketitle

\section{Foreword}
\begin{widetext}
\begin{figure}
	\centering
	\includegraphics[scale=0.48]{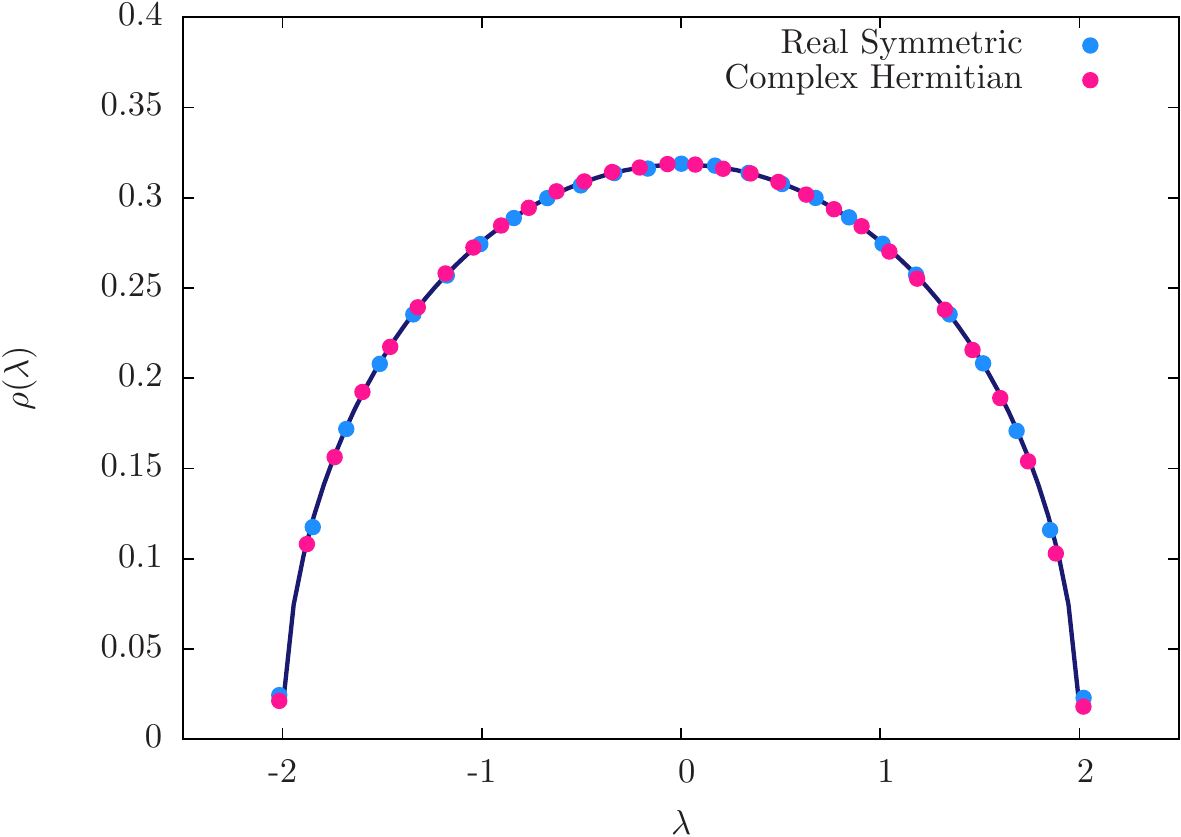}
	\includegraphics[scale=0.48]{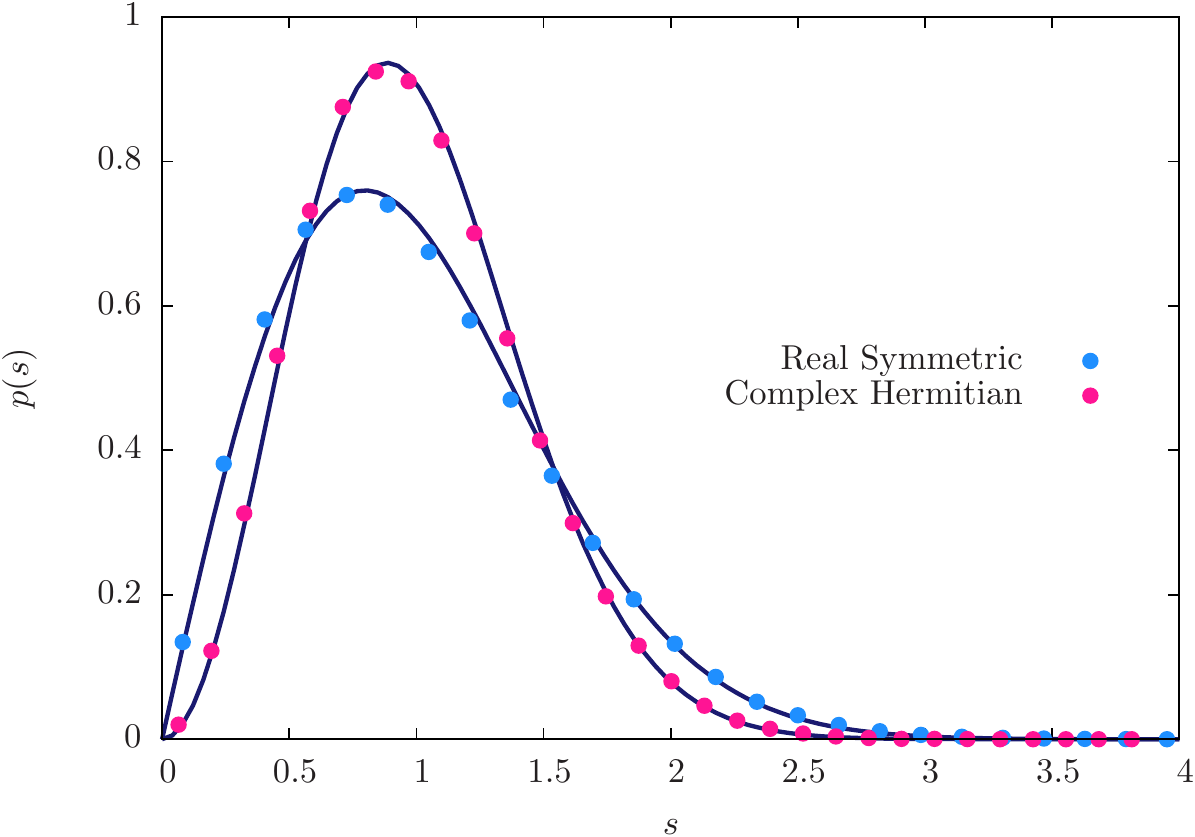}
	\includegraphics[scale=0.48]{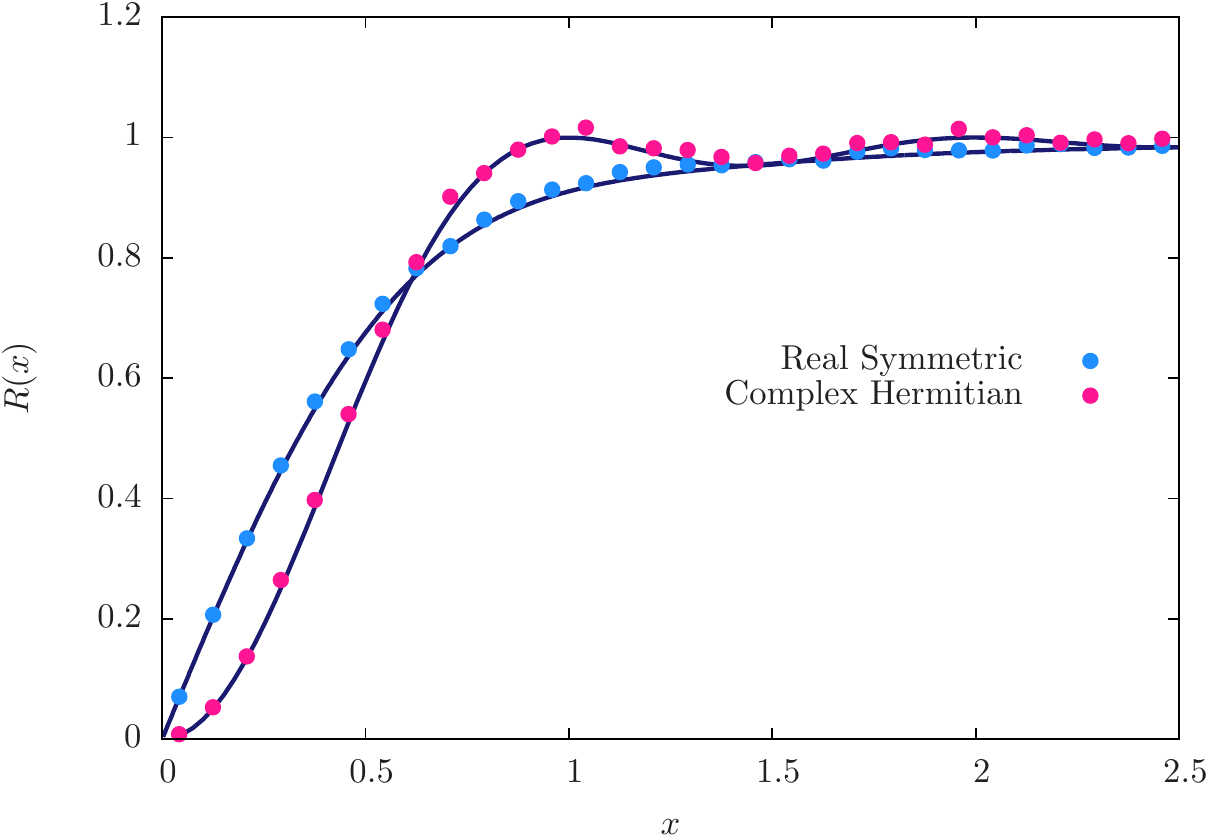}
	\caption{Numerical points: Spectral density (left), level spacing distribution (center) and two-point correlation function (right) for a numerically generated model of real symmetric (complex hermitian) $500 \times 500$ random matrices. In all plots, solid lines refer instead to the large $N$ results for the corresponding quantities of the GOE and GUE.}
	\label{density_spacing_corr}
\end{figure}
\end{widetext}

Consider the following three plots in Fig. \ref{density_spacing_corr}. The numerical points represent (from left to right) the average spectral density, the nearest-neighbor spacing distribution and the two-point correlation function of (unfolded) eigenvalues for a numerically generated ensemble of large random matrices with unitary and orthogonal invariance.

We challenge the reader to guess what this ensemble is, based on the figures at hand.

\section{Introduction}
Since its inception in nuclear physics more than sixty years ago, and much earlier in statistics, Random Matrix Theory (RMT) has become an invaluable tool in many fields of physics and mathematics, with countless applications (see \cite{mehta,forrester,akemann,guhr}). Traditionally, among the ensembles of random matrices with real spectra for which an analytical treatment is feasible (to some extent) we may single out:
\begin{itemize}
\item Matrices with independent identically distributed (i.i.d.) entries (Wigner matrices) \cite{wigner}, such as adjacency matrices of Erd\"os-R\'enyi graphs \cite{reimer}.
\item Matrices with rotational invariance, where the joint probability density (jpd) of the entries $\mathcal{P}[\mathbf{H}]\equiv \mathcal{P}(H_{11},\ldots,H_{NN})$ remains unchanged if one performs a similarity transformation $\mathbf{H}\to \mathbf{O}\mathbf{H}\mathbf{O}^\mathrm{T}$ (for real symmetric matrices $\mathbf{H}$, $\mathbf{O}$ being an arbitrary orthogonal matrix), $\mathbf{H}\to \mathbf{U}\mathbf{H}\mathbf{U}^\dagger$ (for complex hermitian matrices $\mathbf{H}$, $\mathbf{U}$ being an arbitrary unitary matrix) or $\mathbf{H}\to \bm{\Sigma}\mathbf{H}\bm{\Sigma}^*$ (for quaternion self-dual matrices $\mathbf{H}$, $\bm{\Sigma}$ being an arbitrary symplectic matrix). Here, $(\cdot)^{\mathrm{T}}$, $(\cdot)^{\dagger}$ and $(\cdot)^{*}$ stand for transpose, hermitian conjugate and symplectic conjugate of the matrix respectively.

For such matrices, the jpd of eigenvalues $\mathcal{P}(\lambda_1,\ldots,\lambda_N)$ can be generically written as
\begin{equation}
\mathcal{P}(\lambda_1,\ldots,\lambda_N)\propto \mathrm{e}^{-\beta N\sum_{i=1}^N V(\lambda_i)} |\Delta(\bm\lambda)|^\beta\ ,\label{jpd}
\end{equation}
where $\Delta(\bm\lambda)=\prod_{j<k} (\lambda_j-\lambda_k)$ is the Vandermonde determinant, $V(x)$ is a potential suitably growing at infinity, and the Dyson index reads $\beta=1,2,4$ for real symmetric, complex hermitian and quaternion self-dual matrices respectively \footnote{We will not consider the quaternion case ($\beta=4$) here.}. In this case, the eigenvectors are uniformly distributed (with Haar measure) in the corresponding symmetry group (orthogonal, unitary or symplectic respectively).
\end{itemize}
The only ensemble with independent entries and rotational invariance is the Gaussian ensemble $(V(x)=x^2/2)$, where the entries in the upper triangle are independently sampled from a Gaussian distribution (in the real, complex or quaternion domain). These ensembles are then called Gaussian Orthogonal Ensemble (GOE), Gaussian Unitary Ensemble (GUE) and Gaussian Symplectic Ensemble (GSE) respectively.

The presence of the Vandermonde determinant in \eqref{jpd} implies that the eigenvalues of invariant matrix models are strongly correlated variables, whose statistics is very different from the i.i.d. case.  In particular, the distribution of spacings $p(s)$ between adjacent eigenvalues generally develops a behavior like $p(s)\sim s^\beta$ for $s\to 0$, which is known as \emph{level repulsion}. Once the overall density of eigenvalues is discounted from numerically diagonalized ensembles through a procedure called \emph{unfolding}, this repulsion is largely independent of the particular choice of the confining potential, one of the very many manifestations of \emph{universality} in RMT. For general $N$, the spacing distribution has a complicated expression \cite{mehta}, which is however fairly well approximated by an exact calculation for $N=2$, the so called Wigner's surmise
\be
p_{WS}^{(\beta)}(s) = a_\beta s^\beta \mathrm{e}^{-b_\beta s^2} \ ,\label{WS}
\ee
where $a_\beta = 2 \ \Gamma^{\beta+1}((\beta+2)/2) / \Gamma^{\beta+2}((\beta+1)/2)$, $b_\beta = \Gamma^2((\beta+2)/2) / \Gamma^2((\beta+1)/2)$, and $\beta = 1,2,4$ for real symmetric, complex hermitian, and quaternion self-dual matrices respectively.

The tendency of eigenvalues to repel each other is markedly different from the case of uncorrelated random variables on an interval, which tend to cluster and develop a spacing distribution of the Poisson form $p(s)\sim \mathrm{e}^{-s}$. In the field of quantum chaos, such different statistics for the spacings between energy levels helps discriminating between quantum systems whose classical counterpart is chaotic or integrable \cite{bohi}. It is therefore one of the central observables in the so-called \emph{microscopic} analysis of  spectra. 

We consider here a rotationally invariant ensemble of real symmetric or complex hermitian $N\times N$ matrices $\mathbf{S}$ of the form 
\begin{equation}
\mathbf{S}=\sum_{i=1}^M \langle \mathbf{O}_i \mathbf{D}_i\mathbf{O}_i^{\mathrm{T}}\rangle\ ,\qquad \mathbf{S}=\sum_{i=1}^M \langle \mathbf{U}_i \mathbf{D}_i\mathbf{U}_i^\dagger\rangle \ ,\label{defS}
\end{equation}
respectively.
The diagonal matrices $\mathbf{D}_i=\mathrm{diag}\{\lambda_1^{(i)},\ldots,\lambda_N^{(i)}\}$ are built out of independent and identically distributed real eigenvalues drawn from distributions $p^{(i)}(x)$, while the matrices $\mathbf{O}_i$ and $\mathbf{U}_i$ are (uncorrelated) orthogonal or unitary matrices (respectively). The average $\langle\cdot\rangle$ is simultaneously performed over the symmetry group (with Haar measure) and the joint distribution of $\{\lambda_j^{(i)}\}$.

Is it possible to give a more physical interpretation of this model? We can indeed relate it by analogy to the problem of random walk of $N$-dimensional vectors. There are two natural formulations of 
random walk for $N$-dimensional vectors: the first assumes that each component 
of the vector performs an independent random walk, while the second assumes
the radial part (the length) of random walk increments to be a random variable, 
with the angular part being uniformly distributed on the $N$-sphere. In other words,
the length of the increments is a random variable, but the direction of the increments is
uniformly distributed. The second type can be called \emph{radial random walk}, and the two types
are known to belong to the same universality class. We can analogically formulate two types of random walk for matrices:
the first type, where each matrix element performs an
independent random walk was introduced by Dyson. Here, we show that our model $\mathbf{S}$ realizes the counterpart of the radial random walk for matrices.

Indeed, we prescribe a jpd for eigenvalues of individual elements
and rotate the increments randomly using the Haar measure (which implements ``uniformity" over the angular degrees of freedom). 
Obviously, the problem for matrices is more complex than for vectors, since the radial part (given by the set of eigenvalues) may in principle have a complicated jpd. Assuming that the eigenvalues are independent, though, we can show (at least numerically) that for large $N$ the two-point correlators (and in general local statistics) reproduce those known from Dysonian random walk (GOE/GUE). The finite-$N$ details of the two approaches are however entirely different - in particular, the repulsion is not of Vandermondian type at short distances.

Coming back to our model, its rotational invariance is immediate to prove, and in Section \ref{appA} we compute analytically the jpd of eigenvalues $\mathcal{P}(\bm\nu) \equiv\mathcal{P}(\nu_1,\ldots,\nu_N)$ for this model in the complex hermitian (unitary) case as
\be
\mathcal{P}(\bm\nu) \propto \Delta(\bm\nu)\Bigg\langle\int\prod_{j=1}^N \de t_j\frac{\det(\mathrm{e}^{-\mathrm{i}\nu_j t_k})_{j,k=1\to N}}{[\Delta(\bm{t})]^{M-1}}\prod_{i=1}^M\frac{\det(\mathrm{e}^{\mathrm{i}\lambda_j^{(i)}t_k})_{j,k=1\to N}}{\Delta(\bm\lambda^{(i)})}\Bigg\rangle_{\{p^{(i)}(\lambda)\}}\ , \label{jpdeigfinaltext}
\ee
which has a quite unusual ``multi-orthogonal" form (the average here $\langle\cdot\rangle_{\{p^{(i)}(\lambda)\}}$ stands for $\int \prod_{j=1}^N\prod_{i=1}^M \de\lambda_j^{(i)}p^{(i)}(\lambda_j^{(i)})$). If in addition the distribution of eigenvalues is the same for all summands, $p^{(i)}(x)\equiv p(x),\quad\forall i=1,\ldots,M$, Eq. \eqref{jpdeigfinaltext} simplifies to
\begin{align}
\mathcal{P}(\bm\nu) & \propto \Delta(\bm\nu)\int\prod_{j=1}^N \de t_j \frac{\det(\mathrm{e}^{-\mathrm{i}\nu_j t_k})_{j,k=1\to N}}{[\Delta(\bm{t})]^{M-1}}\left[\int\prod_{j=1}^N\de\lambda_j p(\lambda_j)\frac{\det(\mathrm{e}^{\mathrm{i}\lambda_j t_k})_{j,k=1\to N}}{\Delta(\bm\lambda)}\right]^M\ .\label{jpdeigfinalallequal}
\end{align}

Note that the case $M=1$ is trivial, as it corresponds to a simple ``averaged" similarity transformation (so the spectra of $\mathbf{S}$ and $\mathbf{D}_1$ are identical, see also Section \ref{appA}). 

We focus here on the universal features of our model for i.) $N\to\infty$ (for any number $M$ of summands), and ii.) $M\to\infty$ at $N=2$.

\subsection{$N\to\infty$}
In this limit, we numerically find (for several choices of $p(x)$) that classical RMT universality is restored \emph{no matter how many matrices are summed}, i.e. for $M$ as small as $M=2$ (see Fig. \ref{density_spacing_corr}, where we used $N=500$ and $M=2$). This means that local properties (such as the spacing distribution and two-point correlators) are conjectured to converge to the classical (universal) results, even though i.) at the level of individual summands the eigenvalues are completely uncorrelated, and ii.) the interaction between eigenvalues is not Vandermondian \emph{for any size $N$ of the summands} (see \eqref{jpdeigfinaltext}). A proof of this conjecture from \eqref{jpdeigfinaltext} (and the precise conditions on $p(x)$ and its moments) are at present elusive, but this would be a very interesting direction for future research.

The average spectral density of the model (non universal) is instead fully determined by the distributions $p^{(i)}(x)$. The tools provided by \emph{free probability} are particularly suited to the calculation of the spectral density of $\mathbf{S}$ from the ``spectral densities" of the individual summands. For example, in the Foreword we have constructed our ensemble as $\mathbf{D}_1 + \mathbf{O} \mathbf{D}_2 \mathbf{O}^\mathrm{T}$ ($\mathbf{D}_1 + \mathbf{U} \mathbf{D}_2 \mathbf{U}^\dagger$), where $\mathbf{D}_{1,2}$ are diagonal matrices filled with independent and identically distributed elements drawn from a semicircle distribution
\be
p^{(i)}(x)=\frac{1}{2\pi}\sqrt{4-x^2}\quad\forall i\ ,
\ee
whereas $\mathbf{O}$ ($\mathbf{U}$) is a random Haar orthogonal (unitary) matrix\footnote{In order to generate random Haar orthogonal and unitary matrices we employed the algorithm described in \cite{mezzadri}.}. Not surprisingly, the average spectral density of $\mathbf{S}$ is again, of course, the semicircle distribution (see Fig. \ref{density_spacing_corr}), as the semicircle is \emph{stable} under matrix addition. In more general situations, the full formalism of $R$-transforms and Blue functions needs to be employed to compute the spectral density of $\mathbf{S}$.

Note, however, that our result \eqref{jpdeigfinaltext} gives also in principle access to the density of eigenvalues \emph{at finite $N$} and to higher-order correlation functions (not attainable via free probability techniques, see below for more details).

In summary, if we trust the numerical evidence, we have identified two \emph{different} ensembles (our $\mathbf{S}$ and the GOE/GUE) that share the following features, i) they are both rotationally invariant, ii.) they have the same (macroscopic) spectral density for $N\to\infty$, and iii.) they have the same level spacing distribution and two-point correlation function (microscopic) for $N\to\infty$. With the exception of hermitian ensembles \emph{vs.} their fixed-trace counterparts \cite{akem1,akem2}, such correspondences are very unfrequent, and this motivates the challenge we proposed in the Foreword. Note, however, that (at odds with the Gaussian case) the edges of the semicircle are \emph{hard} in our model: even for finite $N$, it is by construction impossible to sample eigenvalues exceeding the edges.

\subsection{$M\to\infty$ for $N=2$}

In the limit of a large number of summands, we find that the distribution $p(s)$ of the spacing $s$ between the two eigenvalues attains a scaling form (see Eq. \eqref{scaling4}), which is however essentially different in the two cases (orthogonal or unitary). In the orthogonal case, we recover the $\beta=1$ Wigner's surmise, while for the unitary case we surprisingly find an entirely new distribution (see Eq. \eqref{scaling2}). Both results are universal, as they only depend on the existence of the variance of the distribution $p(x)$ for eigenvalues of individual summands.

Using the jpd of eigenvalues, we can also compute explicitly the spacing distributions $p(s)$ of $2\times 2$ matrices $\mathbf{S}$ for a finite number of summands, which are instead non-universal (dependent on the distributions $p^{(i)}(x)$ of eigenvalues of individual summands). However, it is interesting to notice that the behavior at $s\to 0^+$ at fixed number $M$ of summands is instead universal and different from the usual RMT benchmark. These calculations are detailed in Appendix \ref{spacing}. In Table \ref{table}, we summarize the interplay between $N$ and $M$ and our findings in each situation.

\begin{table}
\begin{tabular}{ |p{2cm}|p{6.9cm}|p{3cm}|  }
\hline
\multicolumn{3}{|c|}{Summary} \\
\hline
& $N=2$ & $N\gg1$\\
\hline
$M=2$ & $p(s)\approx -s\ln s$ (\underline{Orthogonal}) \eqref{genasy}\newline $p(s)\approx s$ (\underline{Unitary}) \eqref{genasyunitary}\newline (see Appendix \ref{spacing}) & Standard Wigner-Dyson universality (See Fig. \ref{density_spacing_corr} for semicircle law) \\
\hline
$M > 2$ & \underline{Unitary, Gaussian law} \newline $M = 3$: $p(s) \approx -s^2 \ln s$ \eqref{ps3_asy} \newline $M \geq 4$: $p(s) \approx s^2$ \eqref{psM_asy} \newline \underline{Unitary and Orthogonal cases} \newline $M\to\infty$: scaling forms for $p(s)$ \eqref{scaling1},\eqref{scaling2}&
Same behavior as in Fig. \ref{density_spacing_corr} (not shown) \\
\hline
\end{tabular}
\caption{Schematic summary of the interplay between $N$ (size of the matrices) and $M$ (number of summands). For $N=M=2$, the behavior of the spacing distribution for $s\to 0^+$ can be worked out in full generality and shown to be \emph{universal} (irrespective of the distributions $p^{(i)}(x)$ (see Appendix \ref{spacing})). For $N\to\infty$, we investigated numerically the semicircle law and found that classical RMT features are exactly recovered (see Fig. \ref{density_spacing_corr}), irrespective of the number of summands ($M\geq 2$). We conjecture that RMT universality (independence on the distributions $p^{(i)}(x)$) indeed holds for microscopic spectral properties as $N\to\infty$. Finally, for $N=2$ in the unitary case and for a Gaussian distribution of eigenvalues, we find a different behavior of the spacing distribution for $M = 3$ and $M \geq 4$, as shown in the Table. Finally, for $N=2$ and $M\to\infty$, we can invoke a multidimensional version of the Central Limit Theorem to show that the spacing distribution attains scaling forms in the orthogonal and unitary cases (see Section \ref{appvectors}), irrespective of the distribution of diagonal matrices.}\label{table}
\end{table}

\vspace{10pt}
Our paper is a first attempt at a systematic investigation of ensembles of matrices that become asymptotically \emph{free}. The framework of \emph{free probability} is a powerful extension of the concept of independence for random variables to non-commutative objects. Pioneered by Voiculescu \cite{voiculescu}, it has found natural applications in the field of random matrices \cite{speicher1994} where it allows to compute the average density of eigenvalues (in the large matrix size limit $N\to\infty$) for sums or products of random matrices enjoying a property called \emph{freeness}. While its precise mathematical definition is complicated \cite{speicher}, it can be roughly identified with the simultaneous occurrence of the following features for the matrices being summed or multiplied, i.) independence of different matrices, ii.) absence of  correlations between the eigenspaces of different matrices, and iii.) asymptotic limit of large matrix size, $N\to\infty$. If these conditions are satisfied, a general procedure exists to compute the average spectral density of sums (or products) of free random matrices of infinite size, starting from the spectral densities of the individual summands (or factors) (see \cite{burdareview} for a recent review). However, more detailed spectral information (encoded for instance in the jpd of eigenvalues, or the microscopic statistics) is generally unavailable, as well as finite $N$ results of any sort - with the exception of very recent developments for the products of Ginibre and Wishart matrices \cite{akemann1,akemann2}. In this respect, our model offers a unified (and otherwise unavailable) perspective on the microscopic properties of sums of random matrices that become asymptotically free, i.e. the onset of level repulsion starting from uncorrelated constituents and new universal features.

The plan of this paper is as follows. In Section \ref{appA}, we compute the jpd of eigenvalues for our model in the complex hermitian (unitary invariant) case. Then in Section \ref{appvectors} we consider the spacing distribution for $N=2$ and $M\to\infty$, obtaining a new universal limit for the unitary case. This result is obtained via an application of the Multidimensional Central Limit Theorem. In Section \ref{conclusions} we offer concluding remarks, while the Appendix \ref{spacing} is devoted to the spacing distributions for $N=2$ and $M$ small. While such distributions are in general non-universal (dependent on the details of the distribution of eigenvalues of $\mathbf{D}_i$), using two different methods we still find \emph{universal} (but non-Wigner!) behaviors for the spacing distribution as $s\to 0^+$.

\section{The jpd of eigenvalues for the complex hermitian (unitary) case}\label{appA}

The joint distribution of the entries of $\mathbf{S}$ can be written as

\begin{equation}
\mathcal{P}[\mathbf{S}]=\int\prod_{i=1}^M \left[\de\mathbf{U}_i\prod_{j=1}^N \de\lambda_j^{(i)}p^{(i)}(\lambda_j^{(i)})\right]\delta\left(\mathbf{S}-\sum_{i=1}^M \mathbf{U}_i \mathbf{D}_i\mathbf{U}_i^\dagger \right)\ ,
\end{equation}
where $\delta(\cdot)$ is a product of delta function, one for each independent entry of $\mathbf{S}$, and the first integrals are to be performed with a uniform Haar measure on the unitary group. We can now use the following integral representation of the matrix delta function 
\begin{equation}
\delta(\mathbf{H})=\frac{1}{2^N\pi^{N^2}}\int\de\mathbf{T}\ \mathrm{e}^{\mathrm{i}\mathrm{Tr}(\mathbf{TH})}\ ,
\end{equation}
where $\mathbf{T}$ is a $N\times N$ hermitian matrix, to write

\begin{equation}
\mathcal{P}[\mathbf{S}]=\frac{1}{2^N\pi^{N^2}}\int\de\mathbf{T}\ \mathrm{e}^{\mathrm{-i}\mathrm{Tr}(\mathbf{TS})}\int\prod_{i=1}^M \left[\de\mathbf{U}_i\prod_{j=1}^N \de\lambda_j^{(i)}p^{(i)}(\lambda_j^{(i)})\right]\mathrm{e}^{\mathrm{i}\sum_{i=1}^M\mathrm{Tr}\left(\mathbf{T}\mathbf{U}_i \mathbf{D}_i \mathbf{U}_i^\dagger\right)}\ .
\end{equation}

Next, we can use the Harish-Chandra-Itzykson-Zuber (HCIZ) integral formula \cite{harish} to compute the $\de\mathbf{U}_i$ integrals
\begin{equation}
\mathcal{P}[\mathbf{S}]\propto\int\de\mathbf{T}\ \mathrm{e}^{\mathrm{-i}\mathrm{Tr}(\mathbf{TS})}\int\left[\prod_{j=1}^N\prod_{i=1}^M \de\lambda_j^{(i)}p^{(i)}(\lambda_j^{(i)})\right]\prod_{i=1}^M\frac{\det(\mathrm{e}^{\mathrm{i}\lambda_j^{(i)}t_k})_{j,k=1\to N}}{\Delta(\bm\lambda^{(i)})\Delta(\bm{t})}\ .
\end{equation}
We can now diagonalize the hermitian matrix $\mathbf{T}$ by a unitary transformation $\mathbf{T}=\mathbf{WYW^\dagger}$, with $\mathbf{Y}$ the diagonal matrix of eigenvalues $\bm{t}=\{t_1,\ldots,t_N\}$ of $\mathbf{T}$. In a standard way, we get
\begin{align}
\mathcal{P}[\mathbf{S}] \propto\int\prod_{j=1}^N \de t_j\Delta^2(\bm t)\int\de\mathbf{W} \mathrm{e}^{\mathrm{-i}\mathrm{Tr}(\mathbf{WYW^\dagger S})} \int\left[\prod_{j=1}^N\prod_{i=1}^M \de\lambda_j^{(i)}p^{(i)}(\lambda_j^{(i)})\right]\prod_{i=1}^M\frac{\det(\mathrm{e}^{\mathrm{i}\lambda_j^{(i)}t_k})_{j,k=1\to N}}{\Delta(\bm\lambda^{(i)})\Delta(\bm{t})}\ .\label{pinv1}
\end{align}

We can now compute, for a given (arbitrary) unitary matrix $\mathbf{V}$,
\begin{align}
\mathcal{P}[\mathbf{VSV^\dagger}] \propto\int\prod_{j=1}^N \de t_j\Delta^2(\bm t)\int\de\mathbf{W} \mathrm{e}^{\mathrm{-i}\mathrm{Tr}(\mathbf{WYW^\dagger VSV^\dagger})} \int\left[\prod_{j=1}^N\prod_{i=1}^M \de\lambda_j^{(i)}p^{(i)}(\lambda_j^{(i)})\right]\prod_{i=1}^M\frac{\det(\mathrm{e}^{\mathrm{i}\lambda_j^{(i)}t_k})_{j,k=1\to N}}{\Delta(\bm\lambda^{(i)})\Delta(\bm{t})}\ .\label{pinv}
\end{align}
Using the cyclic property of the trace, renaming $\mathbf{V}^\dagger\mathbf{W}\to\mathbf{W}$ and using invariance of the Haar measure, we indeed obtain that \eqref{pinv} is equivalent to \eqref{pinv1}, confirming that $\mathcal{P}[\mathbf{S}]=\mathcal{P}[\mathbf{VSV^\dagger}]$.

Using now the change of variables
\be
\mathcal{P}[\mathbf{S}]\de S_{11}\cdots \de S_{NN}=\mathcal{P}(\bm\nu) \de\nu_1\cdots \de\nu_N\prod_{i,j}\de V_{ij}\ ,
\ee
where $\bm\nu=\{\nu_1,\ldots,\nu_N\}$ are the eigenvalues of $\mathbf{S}$, $\mathcal{P}(\bm\nu) $ their jpd and $V_{ij}$ the independent components of the eigenvectors of $\mathbf{S}$, we obtain for the jpd of eigenvalues of $\mathbf{S}$
\begin{equation}
\mathcal{P}(\bm\nu)   \propto \Delta^2(\bm\nu)\int\prod_{j=1}^N \de t_j\Delta^2(\bm t)\int\de\mathbf{V}\de\mathbf{W} \mathrm{e}^{\mathrm{-i}\mathrm{Tr}(\mathbf{WYW^\dagger VNV^\dagger})}\int\left[\prod_{j=1}^N\prod_{i=1}^M \de\lambda_j^{(i)}p^{(i)}(\lambda_j^{(i)})\right]\prod_{i=1}^M\frac{\det(\mathrm{e}^{\mathrm{i}\lambda_j^{(i)}t_k})_{j,k=1\to N}}{\Delta(\bm\lambda^{(i)})\Delta(\bm{t})}\ ,
\end{equation}
where $\mathbf{N}=\mathrm{diag}(\nu_1,\ldots,\nu_N)$. Using again the cyclic property of the trace, and the invariance of the Haar measure, we can perform another HCIZ integral to get to
\begin{equation}
\mathcal{P}(\bm\nu)   \propto \Delta^2(\bm\nu)\int\prod_{j=1}^N \de t_j\Delta^2(\bm t)\frac{\det(\mathrm{e}^{-\mathrm{i}\nu_j t_k})_{j,k=1\to N}}{\Delta(\bm\nu)\Delta(\bm{t})}\int\left[\prod_{j=1}^N\prod_{i=1}^M \de\lambda_j^{(i)}p^{(i)}(\lambda_j^{(i)})\right]\prod_{i=1}^M\frac{\det(\mathrm{e}^{\mathrm{i}\lambda_j^{(i)}t_k})_{j,k=1\to N}}{\Delta(\bm\lambda^{(i)})\Delta(\bm{t})}\ ,
\end{equation}
which after simplifications reduces to the $N(M+1)$-fold integral
\begin{equation}
\mathcal{P}(\bm\nu)   \propto \Delta(\bm\nu)\int\prod_{j=1}^N \de t_j \left[\prod_{j=1}^N\prod_{i=1}^M \de\lambda_j^{(i)}p^{(i)}(\lambda_j^{(i)})\right]\frac{\det(\mathrm{e}^{-\mathrm{i}\nu_j t_k})_{j,k=1\to N}}{[\Delta(\bm{t})]^{M-1}}\prod_{i=1}^M\frac{\det(\mathrm{e}^{\mathrm{i}\lambda_j^{(i)}t_k})_{j,k=1\to N}}{\Delta(\bm\lambda^{(i)})}\ , \label{jpdeigfinal}
\end{equation}
as in \eqref{jpdeigfinaltext}.

In the case where the distribution of eigenvalues is the same for all summands, $p^{(i)}(x)\equiv p(x),\quad\forall i$, Eq. \eqref{jpdeigfinal} simplifies to
\begin{align}
\mathcal{P}(\bm\nu) & \propto \Delta(\bm\nu)\int\prod_{j=1}^N \de t_j \frac{\det(\mathrm{e}^{-\mathrm{i}\nu_j t_k})_{j,k=1\to N}}{[\Delta(\bm{t})]^{M-1}}\left[\int\prod_{j=1}^N\de\lambda_j p(\lambda_j)\frac{\det(\mathrm{e}^{\mathrm{i}\lambda_j t_k})_{j,k=1\to N}}{\Delta(\bm\lambda)}\right]^M\ .\label{jpdeigfinalallequalsec}
\end{align}

For $M=1$ (no sums), we expect that the jpd of eigenvalues $\bm\nu$ reproduces the jpd of eigenvalues $\bm\lambda$, i.e.
\be
\mathcal{P}(\bm\nu) =\prod_{j=1}^N p^{(1)}(\nu_j)\ ,\label{prodp}
\ee
as the ensemble of matrices $\mathbf{S}$ just contains in this case randomly rotated (and therefore similar) diagonal matrices with identical distribution of elements.
Setting $M=1$ in \eqref{jpdeigfinal} or \eqref{jpdeigfinalallequal}, we have first to evaluate
\be
\int_{-\infty}^\infty\prod_{j=1}^N \de t_j\det(\mathrm{e}^{-\mathrm{i}\nu_j t_k})_{j,k=1\to N}\det(\mathrm{e}^{\mathrm{i}\lambda_j^{(1)} t_k})_{j,k=1\to N}\propto \det\left(\delta(\lambda_j^{(1)}-\nu_k)\right)_{j,k=1\to N}\ ,\label{andredelta}
\ee
where we have used the standard Andr\'eief identity \cite{andre}. Inserting \eqref{andredelta} into \eqref{jpdeigfinal} and expanding the determinant of delta functions, we precisely obtain \eqref{prodp} after making a suitable number of sign changes in the denominator $\Delta(\bm\lambda^{(1)})$ (with $\lambda_j^{(1)}$ replaced by $\nu_k$).

\section{Spacing between the \boldmath{$N=2$} eigenvalues for \boldmath{$M\to\infty$}: interpretation in terms of sums of random vectors}\label{appvectors}
 
In this Section, we discuss the asymptotic properties of the spacing distribution $p(s)$ of the invariant sum \eqref{defS} for $N=2$ and $M\gg 1$. Hereafter we will assume that the eigenvalues of the diagonal matrices $\mathbf{D}_i$ are all drawn from the same distribution with finite variance $\sigma^2$. We are able to  show that $p(s)$ attains for $M\gg 1$ a scaling form
\begin{equation}\label{scaling4}
p(s)=\frac{1}{\sigma\sqrt{M}}\Phi_{\mathbf{O}}\left(\frac{s}{\sigma\sqrt{M}}\right),\qquad p(s)=\frac{1}{\sigma\sqrt{M}}\Phi_{\mathbf{U}}\left(\frac{s}{\sigma\sqrt{M}}\right)
\end{equation}
for the orthogonal and unitary cases respectively. The scaling functions $\Phi_{\mathbf{O}}(x)$ and $\Phi_{\mathbf{U}}(x)$ can be calculated exactly using a quite elegant interpretation of the spacing distribution $p(s)$ in terms of sums of random vectors in a plane (orthogonal case) or in the full three-dimensional space (unitary case). We start by recalling the statement of the Central Limit Theorem for the multidimensional (vectorial) case.

\subsection{Multidimensional Central Limit Theorem}

Let $\{ \mathbf{x}_1,\ldots,\mathbf{x}_M\}$ be a collection of $M$ independent and identically distributed\footnote{The concepts of independence and identical distribution apply to any two different vectors. Within a single vector, its components may well be correlated or non-identically distributed.} vectors in $\mathbb{R}^k$ with mean vector $\bm{\mu}$ and covariance matrix $\bm\Sigma$ (amongst the individual components of the vectors). $\bm\Sigma$ is therefore a $k\times k$ symmetric and positive-semidefinite matrix.

Let
\begin{equation}
\mathbf{x}_i= \left( \begin{array}{c} x_{i(1)} \\ \vdots \\ x_{i(k)}  \end{array} \right) 
\end{equation}
be the $i$th vector, and let us define the sum $\mathbf{s}_M=\sum_{i=1}^M\mathbf{x}_i$. Then the multidimensional Central Limit Theorem states that the normalized sum $\mathbf{s}=(\mathbf{s}_M-M\bm\mu)/\sqrt{M}$ converges for $M\to\infty$ to a multivariate normal (Gaussian) distribution
\begin{equation}
\mathbf{s}\stackrel{D}{\to}\mathcal{N}_k(0,\bm\Sigma)\ .\label{mCLT}
\end{equation}
 Explicitly, the probability density of a multivariate Gaussian variable $\mathbf{s}\sim\mathcal{N}_k(0,\bm\Sigma)$ reads
 \begin{equation}
 \mathcal{P}(\mathbf{s})=\frac{1}{(2\pi)^{k/2}\sqrt{\det\bm\Sigma}}\mathrm{e}^{-\frac{1}{2}\mathbf{s}^{\mathrm{T}}\bm\Sigma^{-1}\mathbf{s}}\ .\label{CLTmulti}
 \end{equation}
 We will be mostly interested in the cases $k=2$ and $k=3$. A natural question is then: what is the distribution of the \emph{length} (the modulus $|\mathbf{s}|$) of the limiting vector sum $\mathbf{s}$ distributed as in \eqref{CLTmulti}? Clearly, an enormous simplification in \eqref{CLTmulti} occurs if the covariance matrix of individual vector components is a multiple of the identity, allowing to use a spherical coordinate transformation. We will compute this distribution for $k=2$ and $k=3$ in due course, after explaining why this machinery turns out to be very useful for our RMT problem.

 \subsection{Spacing of the matrix sum $\mathbf{S}$ as the sum of 2D or 3D random vectors}
 
 \begin{figure}
	\centering
	\includegraphics[scale=1]{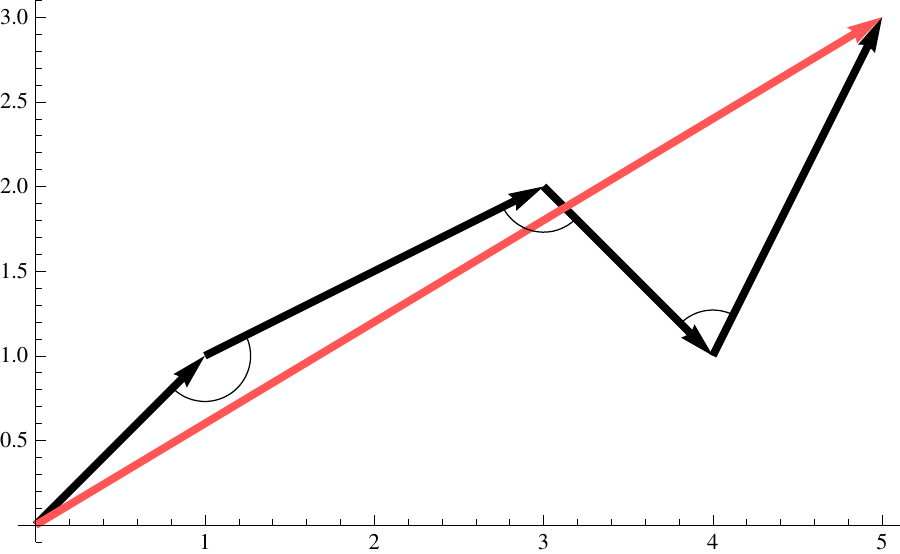}
	\caption{Sketch of the sum of randomly oriented planar vectors.}
	\label{randomvectors}
\end{figure}

Three-dimensional vectors ${\bf x}=(x_1,x_2,x_3)$ can be mapped onto $2\times 2$ matrices ${\bf X} = {\bf x} \cdot \boldsymbol{\sigma} = x_1 \sigma_1 + x_2 \sigma_2 + x_3 \sigma_3$, \emph{i.e.} linear combinations of the Pauli matrices
\begin{align}
\nonumber & \boldsymbol{\sigma}_1 = \left( \begin{array}{rr} 0 & 1 \\ 1 & 0 \end{array} \right) \quad , \qquad
\boldsymbol{\sigma}_2 = \left( \begin{array}{rr} 0 & -\mathrm{i} \\ \mathrm{i} & 0 \end{array} \right) \quad , \qquad
\boldsymbol{\sigma}_3 = \left( \begin{array}{rr} 1 & 0 \\ 0 & -1 \end{array} \right) \quad .\\
\end{align}
Explicitly, we have
\begin{equation}
{\bf X} = \left( \begin{array}{cc} x_3 & x_1-\mathrm{i}x_2 \\ x_1+\mathrm{i}x_2 & -x_3 \end{array} \right)\ .\label{defboldX}
\end{equation}
Note in particular that a vector aligned with the $z$-axis $(x_1=x_2=0)$ corresponds to a \emph{diagonal} matrix ${\bf X}$. The norm of a vector is then given by $|{\bf x}|^2 = - \det{{\bf X}}$, while rotations of a vector ${\bf x} \rightarrow {\bf x}^\prime = {\bf R}{\bf x}$ by an angle $2\phi$ about the axis determined by a unit vector ${\bf n}$ are given by the formula  
\begin{equation}
{\bf X} \rightarrow {\bf X}^\prime = {\bf U} {\bf X} {\bf U}^\dagger
\end{equation}
where ${\bf X}'={\bf x}^\prime \cdot \boldsymbol{\sigma}$ and ${\bf U} = \mathrm{e}^{-\mathrm{i} \phi {\bf n} \cdot \boldsymbol{\sigma}} = \cos(\phi) \boldsymbol{\sigma}_0- \mathrm{i} \sin(\phi) \boldsymbol{\sigma} \cdot {\bf n}$ is a SU(2) matrix ($\boldsymbol{\sigma}_0$ denotes the $2 \times 2$ identity matrix). Explicitly we get
\begin{equation}
 {\bf U} = 
\left(
\begin{array}{cc}
 \cos (\phi )-\mathrm{i} n_3 \sin (\phi ) & (-\mathrm{i} n_1-n_2) \sin (\phi ) \\
  (-\mathrm{i} n_1+ n_2) \sin (\phi ) & \cos (\phi )+\mathrm{i} n_3 \sin (\phi ) \\
\end{array}
\right)\ .
\end{equation}
When rotations about the second axis ${\bf n}=(0,-1,0)$ are considered, the matrix ${\bf U}$ takes the form of an orthogonal matrix ${\bf U} = {\bf O}$
\begin{equation}
{\bf O} =\cos(\phi) \boldsymbol{\sigma}_0+ \mathrm{i} \sin (\phi) \boldsymbol{\sigma}_2 = \left( \begin{array}{rr} \cos (\phi) & \sin (\phi) \\ -\sin (\phi) & \cos (\phi) \end{array} \right) \ . 
\end{equation}
Let us apply this mapping between vectors and $2\times 2$ matrices to analyze the statistical properties of the invariant sum $\mathbf{S}$, starting with the orthogonal case.

\subsubsection{Orthogonal case}\label{appvectorsO}

In this case, we have
\begin{equation}
\label{S1}
{\bf S} = \sum_{i=1}^M {\bf O}_i {\bf D}_i {\bf O}_i^\mathrm{T} \ ,
\end{equation}
where  
\begin{equation}
{\bf D}_i = \left( \begin{array}{rr} \lambda_{1}^{(i)} & 0 \\ 0 & \lambda_{2}^{(i)} \end{array} \right) \quad , \qquad
{\bf O}_i = \left( \begin{array}{rr} \cos (\phi_i) & \sin (\phi_i) \\ -\sin (\phi_i) & \cos (\phi_i) \end{array} \right) \ ,
\end{equation}
the $\phi_i$'s being independent random variables uniformly distributed over $[0,\pi/2]$, and the $\lambda_{j}^{(i)}$'s are i.i.d. random variables for $i=1,\ldots,M$ and $j=1,2$. In \eqref{ortmatrix}, we used the identification $\phi_{i=1}\equiv t$.
The diagonal matrices ${\bf D}_i$ can be written as
\begin{equation}
\label{diagdecomp}
{\bf D}_i = \frac{t_i}{2} \boldsymbol{\sigma}_0 + \frac{s_i}{2} \boldsymbol{\sigma}_3 \ ,
\end{equation}
where $t_i = \lambda_{1}^{(i)} + \lambda_{2}^{(i)}$ is the trace of ${\bf D}_i$, and $s_i = \lambda_{1}^{(i)} - \lambda_{2}^{(i)}$ is the spacing between the unsorted eigenvalues of $\mathbf{D}_i$ (so $s_i\lessgtr 0$).
From equation \eqref{diagdecomp} we see that the invariant sum \eqref{S1} takes the form
\begin{equation}
{\bf S} = \frac{1}{2} \sum_{i=1}^M t_i \boldsymbol{\sigma}_0 + \frac{1}{2} \sum_{i=1}^M {\bf O}_i (s_i \boldsymbol{\sigma}_3) {\bf O}_i^\mathrm{T} \ . \label{repS}
\end{equation}
The first term shows that the trace of the sum is a sum of traces of individual terms, whereas the second term can be interpreted as a sum of planar vectors embedded in the plane $(x,z)$. Indeed, each term in the sum
\begin{equation}
{\bf O}_i (s_i \boldsymbol{\sigma}_3) {\bf O}_i^\mathrm{T} =  {\bf x}_i^\prime \cdot \boldsymbol{\sigma}= 
- \sin(2\phi_i) s_i \boldsymbol{\sigma}_1 + \cos(2\phi_i) s_i \boldsymbol{\sigma}_3  
\end{equation}
represents a vector ${\bf x}^\prime_i$ of length $s_i$ obtained rotating the vector ${\bf x}_i=(0,0,s_i)$ around the second axis by an angle $2\phi_i$.  The rotated vector 
\begin{equation}
{\bf x}^\prime_i=(-\sin(2\phi_i) s_i,0,\cos(2\phi_i) s_i)\label{xprimedef}
\end{equation}
remains in the plane $(x,z)$ and so does the sum ${\bf s}=\sum_i {\bf x}_i^\prime$, corresponding to the second term in \eqref{repS}, $ \sum_{i=1}^M {\bf O}_i (s_i \boldsymbol{\sigma}_3) {\bf O}_i^\mathrm{T} $. This vector ${\bf s}$ can be rotated back and aligned to the third axis $z$ by a rotation about the second axis. But recall that a vector aligned to the axis $z$ corresponds to a $2\times 2$ \emph{diagonal} matrix, from \eqref{defboldX}. Hence, this rotation of the vector ${\bf s}$ corresponds, in the matrix language, to the diagonalization of $\sum_{i=1}^M {\bf O}_i (s_i \boldsymbol{\sigma}_3) {\bf O}_i^\mathrm{T} $. Using \eqref{repS}, one concludes that the matrix $\mathbf{S}$ itself becomes diagonal, i.e.
\begin{equation}
\left (\sum_{i=1}^M {\bf x}_i^\prime \right ) \cdot \boldsymbol{\sigma} = {\bf O} \left( s \boldsymbol{\sigma}_3\right) {\bf O}^\mathrm{T} \ ,\label{boldOrotate}
\end{equation}
where ${\bf O} = \cos (\phi) \boldsymbol{\sigma}_0 + i \sin(\phi) \boldsymbol{\sigma}_2$ is the orthogonal matrix that diagonalizes $\mathbf{S}$, and $2\phi$ is the
angle between ${\bf s}$ and the $z$-axis. The eigenvalues of $\mathbf{S}$, combining \eqref{repS} and \eqref{boldOrotate}, can be written as $\nu_1=t+s$ and $\nu_2=t-s$, implying that $s$ is indeed the spacing of the invariant sum $\mathbf{S}$. Since rotation does not change the length $s=|{\bf s}|$ of the vector, we see from \eqref{repS} that the spacing (between the \emph{largest} and the \emph{smallest}, i.e. $s>0$) is given by length of the sum ${\bf s}=\sum_i {\bf x}_i^\prime$ of $M$ identically distributed random vectors with random directions in the plane $(x,z)$. For $M\gg 1$, we can then apply the multidimensional Central Limit Theorem as in \eqref{mCLT}: all we need is the average vector $\bm\mu$ and the covariance $\bm\Sigma$ between the components of each summand. 

\begin{figure}
	\centering
	\includegraphics[scale=0.9]{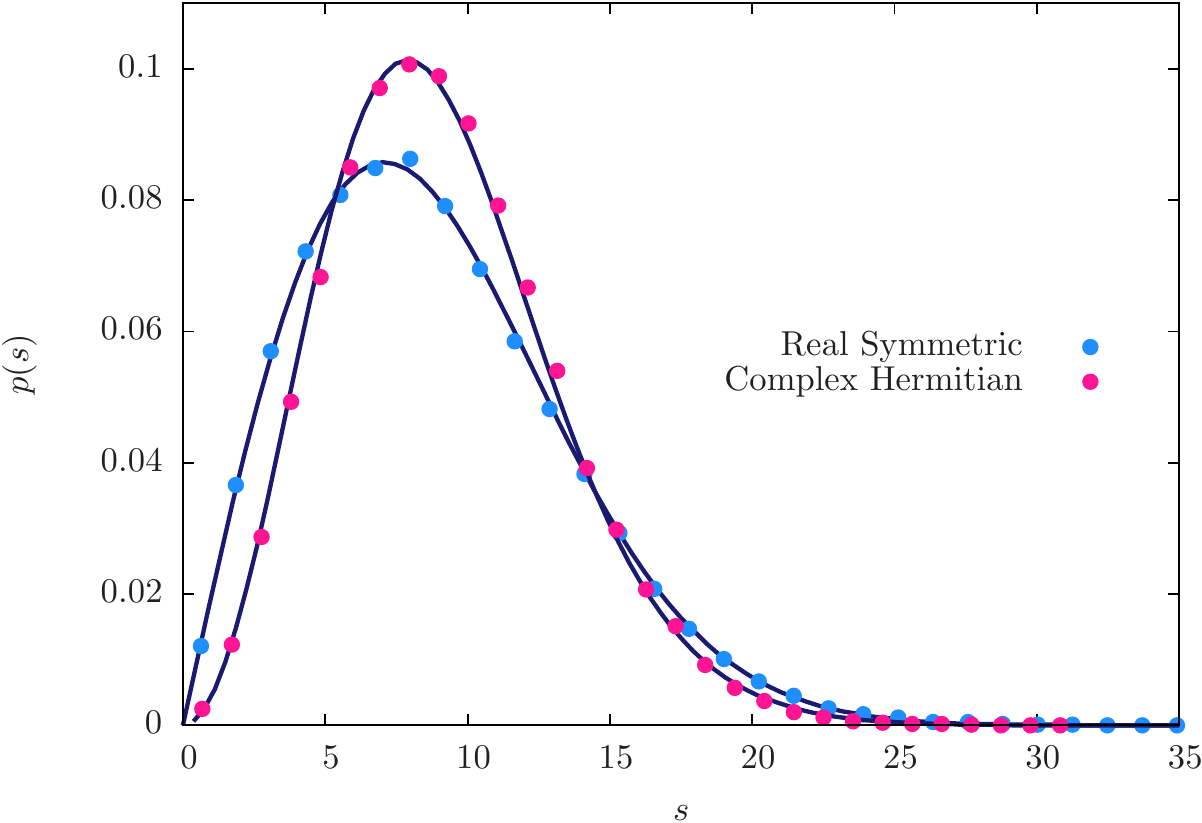}
	\caption{Spacing distribution for $2 \times 2$ real symmetric matrices formed as in equation \eqref{S1} and $2 \times 2$ complex Hermitian matrices formed as in equation \eqref{S3}, where entries in the diagonal matrices ${\bf D}_i$ are i.i.d. random variables drawn from a standard Gaussian distribution. Numerical results (dots) refer to the sum of $M = 50$ matrices, whereas the blue solid lines represent the analytical predictions of equation \eqref{scaling1} (orthogonal case) and \eqref{scaling2} (unitary case) with $\sigma = 1$.}
	\label{spacing_2x2_M_scaling}
\end{figure}

We have from \eqref{xprimedef}
\begin{align}
&\langle\sin(2\phi_i)s_i\rangle =\langle\cos(2\phi_i)s_i\rangle=0\ ,\\
&\langle (\sin(2\phi_i)s_i)^2\rangle =\langle(\cos(2\phi_i)s_i)^2\rangle=\frac{1}{2}\times 2 \sigma^2=\sigma^2\ ,\\
&\langle\sin(2\phi_i)\cos(2\phi_i)s_i^2\rangle =0\ ,
\end{align}
where, since the angles and the components $s_i$ are independent of each other, the average $\langle (\cdots)\rangle$ factorizes into the product of the average over a flat distribution of the angles over $[0,\pi/2]$, and the second moment of the individual spacing $\langle (s_i)^2\rangle = 2 \sigma^2$, if $\sigma^2$ is the variance of the individual eigenvalues $\lambda_{(1,2)}^{(i)}$. Note that we have used the fact that $\langle s_i\rangle=0$, as the $s_i$ can be $\lessgtr 0$ (contrary to the total spacing $s>0$) and we recall the assumption that $\lambda_{1,2}^{(i)}$ have been both drawn from the same distribution with finite variance $\sigma^2$. Hence, the covariance matrix $\bm\Sigma$ is in this case a multiple of the identity,
\begin{equation}
\bm\Sigma=\sigma^2\bm\sigma_0\ .\label{SigmaOrt}
\end{equation}
Taking into account the scaling factor $\sqrt{M}$ that appears on the left hand side of \eqref{mCLT}, we conclude that the distribution of the \emph{length} of the vector $s=|\mathbf{s}|$ (i.e. the spacing of the sum $\mathbf{S}$) can be computed from \eqref{CLTmulti} as
\begin{equation}
p(s)\propto \int_{(-\infty,\infty)^2}\de s_1\de s_2 \mathrm{e}^{-\frac{s_1^2+s_2^2}{2 M\sigma^2}}\delta\left(s-\sqrt{s_1^2+s_2^2}\right)\ ,
\end{equation}
which can be easily computed in polar coordinates, yielding eventually the scaling form for $M\gg 1$
\begin{equation}
p(s)=\frac{1}{\sigma\sqrt{M}}\Phi_{\mathbf{O}}\left(\frac{s}{\sigma\sqrt{M}}\right),\qquad \Phi_{\mathbf{O}}(x)=x\ \mathrm{e}^{-\frac{1}{2}x^2}\ .\label{scaling1}
\end{equation}
Note that the scaling function $\Phi_{\mathbf{O}}(x)$ has precisely the form of the Wigner's surmise $p_{WS}^{(\beta)}(x)$ for $\beta=1$ (see \eqref{WS}), giving the spacing distribution of Gaussian matrices with orthogonal symmetry (see Figure \ref{spacing_2x2_M_scaling} for a numerical validation). Therefore, in the orthogonal case the spacing distributions in the limiting situations $N\gg 1,M=2$ (on unfolded eigenvalues) and $N=2,M\gg 1$ are the same.

\subsubsection{Unitary case}\label{appvectorsU}

We can repeat the same argument for the unitary case
\begin{equation}
\label{S3}
\mathbf{S} = \sum_{i=1}^M \mathbf{U}_i \mathbf{D}_i \mathbf{U}_i^\dagger \ .
\end{equation}
The main difference with respect to the previous case is that the rotations 
\begin{equation}
\mathbf{U}_i (s_i \bm\sigma_3) \mathbf{U}_i^\dagger =  \mathbf{x}_i' \cdot \bm\sigma
\end{equation}
produce vectors $\mathbf{x}^\prime$ which have random directions not in two but in three dimensional space,
\begin{equation}
\mathbf{x}^\prime=(s_i\sin(2t)\cos(\theta+\phi_1+\phi_2),s_i\sin(2 t)\sin(\theta+\phi_1+\phi_2),s_i\cos(2 t))\ ,
\end{equation}
where we have used the parametrization \eqref{paramunitary} for a general $2\times 2$ unitary matrix. We then have
\begin{align}
&\langle s_i\sin(2t)\cos(\theta+\phi_1+\phi_2)\rangle =\langle s_i\sin(2t)\sin(\theta+\phi_1+\phi_2)\rangle=\langle\cos(2 t)\rangle=0\ ,\\
&\langle (s_i\sin(2t)\cos(\theta+\phi_1+\phi_2))^2\rangle =\langle (s_i\sin(2t)\sin(\theta+\phi_1+\phi_2))^2\rangle=\frac{1}{4}\times 2\sigma^2=\frac{1}{2}\sigma^2\ ,\\
&\langle(s_i\cos(2 t))^2\rangle =\frac{1}{2}\times 2\sigma^2=\sigma^2\ ,\\
&\nonumber\langle s_i^2\sin(2t)^2\cos(\theta+\phi_1+\phi_2)\sin(\theta+\phi_1+\phi_2)\rangle=\langle s_i^2\sin(2t)\cos(2 t)\cos(\theta+\phi_1+\phi_2)\rangle=\\
&=\langle s_i^2\sin(2t)\cos(2 t)\sin(\theta+\phi_1+\phi_2)\rangle=0\ .
\end{align}
Hence, in this case the covariance matrix is still diagonal, but is no longer a multiple of the identity
\begin{equation}
\bm\Sigma =\sigma^2\mathrm{diag}\left(\frac{1}{2},\frac{1}{2},1\right)\ ,
\end{equation}
implying that the spacing distribution (invoking again \eqref{mCLT} and \eqref{CLTmulti}) reads in this case
\begin{equation}
 p(s)\propto\int_{(-\infty,\infty)^3}\de s_1\de s_2\de s_3 \exp\left[-\frac{1}{2 M\sigma^2}(2 s_1^2+2 s_2^2+s_3^2)\right]\delta\left(s-\sqrt{s_1^2+s_2^2+s_3^2}\right)\ .
\end{equation}
The integral can be still performed in spherical coordinates, and again a scaling form (valid for $M\gg 1$) is attained
\begin{equation}
p(s)=\frac{1}{\sigma\sqrt{M}}\Phi_{\mathbf{U}}\left(\frac{s}{\sigma\sqrt{M}}\right),\qquad \Phi_{\mathbf{U}}(x)=2 x\ \mathrm{e}^{-x^2}\mathrm{erfi}\left(\frac{x}{\sqrt{2}}\right)\ ,\label{scaling2}
\end{equation}
where $\mathrm{erfi}(z)=-\mathrm{i}\ \mathrm{erf}(\mathrm{i}z)$. The scaling function $\Phi_{\mathbf{U}}(x)$ is correctly normalized $\int_0^\infty\de x\Phi_{\mathbf{U}}(x)=1$, and for $x\to 0^+$ it goes like $\Phi_{\mathbf{U}}(x)\sim 2\sqrt{2/\pi} x^2$. The quadratic behavior is in agreement with the finite-$M$ formulae we derive in Appendix \ref{spacing} (equations \eqref{psM} and \eqref{JMSapp}), which hold for $M\geq 4$. However, it is not easy to derive \eqref{scaling2} directly from \eqref{JMSapp}. This new universal scaling function is instead significantly different from the Wigner's surmise for $\beta=2$ (see \eqref{WS}). In Figure \ref{spacing_2x2_M_scaling} a numerical validation of equation \eqref{scaling2} is shown.

\section{Conclusions}\label{conclusions}
In summary, using a simple ``rotate and sum" procedure that is customary in RMT approaches based on free probability, we computed analytically the jpd of eigenvalues and level spacing statistics for an ensemble of random matrices with rather unusual features. It is by construction invariant with respect to classical symmetry groups (orthogonal and unitary) and yet its eigenvalue repulsion is not precisely Vandermondian. We focused on the universal limits i.) $N\to\infty$ and ii.) $M\to\infty$, with $N=2$, while in Appendix \ref{spacing} we will discuss the spacing distribution for $N=2$ and finite (small) number of summands. In the limit i.), the resulting sum $\mathbf{S}$ develops the classical level repulsion even though a) the original matrices have uncorrelated eigenvalues, and b) the interaction between eigenvalues of $\mathbf{S}$ is of a new (non-Vandermondian) form (see \eqref{jpdeigfinaltext}) for any value of $N$. In the limit ii.) the spacing distribution attains scaling forms that are computed exactly using the multidimensional Central Limit Theorem for the sum of random vectors in the plane (orthogonal case) or in three-dimensional space (unitary case): for the orthogonal case, we recover the $\beta=1$ Wigner's surmise \eqref{WS}, while for the unitary case we discover an entirely new universal distribution. It would indeed be interesting to understand in more detail how the transition from the scaling function $\Phi_{\mathbf{U}}(x)$ \eqref{scaling2} to the Wigner's surmise \eqref{WS} for $\beta=2$ occurs upon increasing the size $N$. Distributions of eigenvalues with infinite variance, belonging to
the L\'evy-Khintchine universality class, may lead to different scaling functions and this constitutes another possible direction for further research \cite{burdaheavy}.

Our simple construction offers analytical access to the microscopic statistics of the sum of random matrices that become asymptotically free, for both (at least in principle) small and large $N$ and $M$, a task that is usually out of the range of standard tools of free probability. All our results have been corroborated by numerical simulations. Our work can stimulate further research in the following directions: on one hand, it would be interesting to analyze in more detail the jpd \eqref{jpdeigfinaltext} of our model and its marginals (average density and correlation functions). On the other hand, further analytical insight into the limit $N\to\infty$ could be perhaps gained by analyzing the large-$N$ limit of the HCIZ integrals appearing in \ref{appA} \cite{majbouchaud}. It will also be interesting to see if more exotic choices for the distributions $p^{(1,2)}(x)$ may break the universality of the spacing distribution for $s\to 0^+$ at fixed $M$ (see Appendix \ref{spacing}). The limit $M\to\infty$ at fixed (small) $N>2$ is also interesting and is left for future investigation. Finally, the recently introduced \emph{spacing ratio distribution} \cite{ratio1,ratio2} will be well worth studying in our model, for example for $N=3$ (the smallest sensible size). This should then be compared with the classical GOE and GUE results, which are known to be highly universal, in particular regarding the role played by the number of summands $M$.\\

\begin{acknowledgments}
We thank Gernot Akemann for useful comments and for pointing out relevant references. 
This work is supported by ``Investissements d'Avenir" LabEx PALM (ANR-10-LABX-0039-PALM) [P.V.] and by the Grant DEC-2011/02/A/ST1/00119 of
the National Centre of Science [Z.B.]. 
\end{acknowledgments}

\appendix 

\section{Level Spacings \boldmath{$p(s)$} for \boldmath{$N=M=2$} }\label{spacing}

In this Appendix, we consider the case $N=M=2$ and the spacing distribution $p(s)$ from two different starting points. First, we compute the spacing distribution directly from the definition of the matrix $\mathbf{S}$ in \eqref{defS} for generic distributions of the eigenvalues of the two summands, and we unveil a universal behavior of $p(s)$ for $s\to 0^+$ which is different from the standard Wigner's surmise. Next, we reproduce the same results (specialized to the unitary case and Gaussian distributed eigenvalues), this time starting from the jpd \eqref{jpdeigfinaltext}. Eventually, starting again from the jpd, we consider the case of generic $M>2$.

\subsection{From the general definition of matrices $\mathbf{S}$}

In this subsection, we compute analytically the probability density $p(s)$ of the level spacing $s=\nu_2-\nu_1$ for the case $N=M=2$, for real symmetric and complex hermitian matrices $\mathbf{S}$. We offer here a derivation uniquely based on the matrix representation \eqref{defS}. This allows to i.) treat the orthogonal and unitary cases on the same footing, and ii.) to cross-check the result for the unitary case against a derivation based on the jpd \eqref{jpdeigfinaltext} (see next subsection). We assume without loss of generality the following form for the matrix sum
\begin{equation}
\mathbf{S}=\mathbf{D}_1+\langle \mathbf{U} \mathbf{D}_2\mathbf{U}^\dagger\rangle\ ,\qquad \mathbf{S}=\mathbf{D}_1+\langle \mathbf{O} \mathbf{D}_2\mathbf{O}^{\mathrm{T}}\rangle\ ,
\end{equation}
where $\mathbf{U}$ is a unitary $2\times 2$ matrix, $\mathbf{O}$ is a orthogonal $2\times 2$ matrix and $\langle\cdot\rangle$ stands for integration over the respective group. For simplicity, we rename the eigenvalues as $(\lambda_1,\lambda_2)$ for the diagonal matrix $\mathbf{D}_1$ and $(\mu_1,\mu_2)$ for $\mathbf{D}_2$. They are independently drawn from probability densities $p^{(1)}(\lambda), p^{(2)}(\mu)$ respectively. The matrix $\mathbf{U}$ can be parametrized as 
\begin{equation}
\mathbf{U}=
\left(
\begin{array}{cc}
 \mathrm{e}^{-\mathrm{i} \phi _1} \cos (t) & -\mathrm{e}^{-\mathrm{i} \theta -\mathrm{i} \phi _1} \sin (t) \\
 \mathrm{e}^{\mathrm{i} \theta +\mathrm{i} \phi _2} \sin (t) & \mathrm{e}^{\mathrm{i} \phi _2} \cos (t) \\
\end{array}
\right)\label{paramunitary}
\end{equation}
for the unitary case, and
\begin{equation}
\mathbf{O}=
\left(
\begin{array}{cc}
\cos(t)& \sin(t) \\
-\sin(t) & \cos(t) \\
\end{array}
\right)\label{ortmatrix}
\end{equation}
for the orthogonal case, where $0\leq \phi_1,\phi_2,\theta\leq 2\pi$ and $0\leq t\leq \pi/2$. It turns out that the spacing $s$ between the largest and the smallest eigenvalue of $\mathbf{S}$ has an expression that depends only on the angle $t$ and is the same for the unitary and orthogonal cases, $s=\sqrt{q}$, where
\begin{equation}
q=\left(\lambda_2-\lambda_1\right)^2+\left(\mu_2-\mu_1\right)^2+2\left(\lambda_2-\lambda_1\right)\left(\mu_2-\mu_1\right)\cos(2t)\ .\label{defq}
\end{equation}
The probability density of $q$ is therefore given by 
\begin{align}
&\nonumber \mathcal{P}(q) =\frac{2}{\pi}\int_{\sigma_1} \de\lambda_1 \de\lambda_2 p^{(1)}(\lambda_1)p^{(1)}(\lambda_2)\int_{\sigma_2} \de\mu_1 \de\mu_2 p^{(2)}(\mu_1)p^{(2)}(\mu_2)\int_0^{\pi/2}\de t\times\\ &\times \delta\left(q-\left[\left(\lambda_2-\lambda_1\right)^2+\left(\mu_2-\mu_1\right)^2+2\left(\lambda_2-\lambda_1\right)\left(\mu_2-\mu_1\right)\cos(2t)\right]\right)\ ,\label{orthopq}
\end{align}
for the orthogonal case, and
\begin{align}
&\nonumber \mathcal{P}(q) =2\int_{\sigma_1} \de\lambda_1 \de\lambda_2 p^{(1)}(\lambda_1)p^{(1)}(\lambda_2)\int_{\sigma_2} \de\mu_1 \de\mu_2 p^{(2)}(\mu_1)p^{(2)}(\mu_2)\int_0^{\pi/2}\de t\cos(t)\sin(t)\times\\
&\times \delta\left(q-\left[\left(\lambda_2-\lambda_1\right)^2+\left(\mu_2-\mu_1\right)^2+2\left(\lambda_2-\lambda_1\right)\left(\mu_2-\mu_1\right)\cos(2t)\right]\right)\ ,\label{unitpq}
\end{align}
for the unitary case, where the first four integrals run over the supports $\sigma_1$ and $\sigma_2$ of each probability density (in general distinct). Note that for $M\geq 3$, instead, the expression for $q$ is different in the orthogonal and unitary cases, but it turns out to have a simple interpretation in terms of sums of $M$ random two-dimensional (orthogonal case) or three-dimensional (unitary case) vectors (see Sec. \ref{appvectors}).

Then the spacing distribution in both cases is given by
\begin{equation}
p(s)=2s\ \mathcal{P}(s^2)\ .\label{qtos}
\end{equation}
In order to get a manageable expression for $\mathcal{P}(q)$, we first introduce the two identities
\begin{align}
1 &=\int_{-\infty}^\infty \de s_1 \delta\left(s_1-\left(\lambda_2-\lambda_1\right)\right)\\
1 &=\int_{-\infty}^\infty \de s_2 \delta\left(s_2-\left(\mu_2-\mu_1\right)\right)\ ,
\end{align}
as well as the following auxiliary functions
\begin{align}
\omega^{(1)}(s) &=\int_{\sigma_1}\de\lambda_1\int_{\sigma_1}\de\lambda_2 \delta(s-(\lambda_2-\lambda_1))p^{(1)}(\lambda_1)p^{(1)}(\lambda_2)\label{pi}\\
\omega^{(2)}(s) &=\int_{\sigma_2}\de\mu_1\int_{\sigma_2}\de\mu_2 \delta(s-(\mu_2-\mu_1)) p^{(2)}(\mu_1)p^{(2)}(\mu_2)\ .\label{pitilde}
\end{align}
Now we consider the two cases (orthogonal and unitary) separately.

\subsubsection{Orthogonal case}
We can now rewrite \eqref{orthopq} as
\begin{equation}
 \mathcal{P}(q) =\frac{2}{\pi}\int_{-\infty}^\infty \de s_1 \de s_2\int_0^{\pi/2}\de t\ \delta\left(q-\left[s_1^2+s_2^2+2 s_1 s_2\cos(2t)\right]\right)\omega^{(1)}(s_1)\omega^{(2)}(s_2)\ .
\end{equation}
Everything is therefore expressed only in terms of the auxiliary functions $\omega^{(1)}(s)$ and $\omega^{(2)}(s)$ which represent a sort of spacing distributions (including sign!) of the individual diagonal matrices $\mathbf{D}_1$ and $\mathbf{D}_2$. To make further progress, it is convenient to introduce the change of variables $s_1=(x-y)/2$ and $s_2=(x+y)/2$, yielding
\begin{align}
\nonumber \mathcal{P} (q) &=\frac{1}{\pi}\int_{-\infty}^\infty \de x \de y\ \omega^{(1)}\left(\frac{x-y}{2}\right)\omega^{(2)}\left(\frac{x+y}{2}\right)\times\\
&\times \int_0^{\pi/2}\de t\ \delta\left(q-\frac{1}{2}(x^2+y^2)-\frac{1}{2}(x^2-y^2)\cos(2 t)\right)\ .
\end{align}
Setting $\cos(2t)=\xi$, one obtains
\begin{align}
\nonumber &\mathcal{P}(q) =\frac{1}{2\pi}\int_{-\infty}^\infty \de x \de y\ \omega^{(1)}\left(\frac{x-y}{2}\right)\omega^{(2)}\left(\frac{x+y}{2}\right)\times \\ &\times \int_{-1}^1 \frac{\de\xi}{\sqrt{1-\xi^2}}\frac{\delta\left(\xi-\frac{q-(1/2)(x^2+y^2)}{(1/2)(x^2-y^2)}\right)\mathds{1}\left(-1\leq \frac{q-(1/2)(x^2+y^2)}{(1/2)(x^2-y^2)}\leq 1\right)}{(1/2)|x^2-y^2|}\ ,
\end{align}
where $\mathds{1}(x)$ is the indicator function, equal to $1$ if $x$ is logically true and $0$ otherwise. Resolving the constraints, we can express $ \mathcal{P}(q)$ as the sum of four contributions
\begin{equation}
 \mathcal{P}(q)=\frac{1}{2\pi}\sum_{j=1}^4 I_j(q),\qquad I_j(q)=\int_{X_j}\de x\int_{Y_j}\de y f_q(x,y)\ ,
\end{equation}
where 
\begin{equation}
f_q(x,y)=\frac{\omega^{(1)}\left(\frac{x-y}{2}\right)\omega^{(2)}\left(\frac{x+y}{2}\right)}{\sqrt{(x^2-q)(q-y^2)}}\label{fq}
\end{equation}
and $X_1=Y_2=(-\infty,-\sqrt{q})$, $X_2=X_3=Y_1=Y_4=(-\sqrt{q},\sqrt{q})$ and $X_4=Y_3=(\sqrt{q},\infty)$. It is easy to see that 
if $\omega^{(1)}(x)=\omega^{(1)}(-x)$, then $I_1(q)=I_2(q)$ and $I_3(q)=I_4(q)$, and if $\omega^{(1)}(x)=\omega^{(2)}(x)$, then $I_2(q)=I_3(q)$.

Therefore, if $\omega^{(1)}(x)=\omega^{(2)}(x)$ is an even function, then $\mathcal{P}(q)$ can be simplified as
\begin{equation}
\mathcal{P}(q)=\frac{1}{2\pi}\times 4 \int_{X_j}\de x\int_{Y_j}\de y f_q(x,y),\qquad \mbox{any }j\ .\label{anyj}
\end{equation}
We shall mainly restrict to this case henceforth. Furthermore if $0<\omega^{(1)}(0)<\infty$, we prove that for $M=N=2$ the level repulsion at zero is universally given by the following non-Wigner behavior 
\begin{equation}
p(s)\sim -4 [\omega^{(1)}(0)]^2 s\ln s\ ,\qquad s\to 0^+\ ,\label{genasy}
\end{equation}
a direct consequence of the fact that the repulsion between eigenvalues of $\mathbf{S}$ is not precisely Vandermondian, as in the standard invariant ensembles. A similar, weaker repulsion of eigenvalues was detected in another random matrix models for pseudo-Hermitian matrices \cite{spacingunc}. In the asymptotic limits $M\geq 2,N\to\infty$, however, the standard Wigner-surmise behavior is recovered after unfolding (see Fig. \ref{density_spacing_corr}, top right, for the case $M=2$).

In order to prove \eqref{genasy}, we start from \eqref{anyj} (combined with \eqref{qtos}) in the form
\begin{equation}
\label{orth_spacing}
p(s)= \frac{4}{\pi} s \int_s^\infty \de x\int_{-s}^s \de y \frac{\omega^{(1)}\left(\frac{x-y}{2}\right)\omega^{(1)}\left(\frac{x+y}{2}\right)}{\sqrt{(x^2-s^2)(s^2-y^2)}}\ .
\end{equation}
For small $s$, the integral in $y$ can be estimated as
\begin{equation}
\int_{-s}^s \de y \frac{\omega^{(1)}\left(\frac{x-y}{2}\right)\omega^{(1)}\left(\frac{x+y}{2}\right)}{\sqrt{s^2-y^2}}\to \left[\omega^{(1)}\left(\frac{x}{2}\right)\right]^2 \int_{-s}^s \frac{\de y}{\sqrt{s^2-y^2}}=\pi  \left[\omega^{(1)}\left(\frac{x}{2}\right)\right]^2\ .
\end{equation}
Therefore
\begin{equation}
p(s)\sim 4s \int_s^\infty \de x \frac{\left[\omega^{(1)}\left(\frac{x}{2}\right)\right]^2}{\sqrt{x^2-s^2}}\ .\label{intx1}
\end{equation}
Any singular behavior of the integral \eqref{intx1} can only arise in the vicinity of $x\simeq s$. Therefore, for $s$ close to zero we can estimate the integral contribution as
\begin{equation}
\int_s^\infty \de x \frac{\left[\omega^{(1)}\left(\frac{x}{2}\right)\right]^2}{\sqrt{x^2-s^2}}\to \left[\omega^{(1)}\left(0\right)\right]^2\int_s^1 \frac{\de x}{\sqrt{x^2-s^2}}=\left[\omega^{(1)}\left(0\right)\right]^2 \ln \left(\frac{1+\sqrt{1-s^2}}{s}\right)\ .
\end{equation}
Expanding the logarithm around $s=0$, and collecting prefactors we precisely arrive at \eqref{genasy}.
We can verify this general statement by explicitly drawing the eigenvalues of the matrices $\mathbf{D}_1$ and $\mathbf{D}_2$ e.g. from a standard normal distribution
\begin{equation}
p^{(1)}(x)=p^{(2)}(x)=\frac{\mathrm{e}^{-\frac{1}{2}x^2}}{\sqrt{2\pi}}\ .
\end{equation}
Performing the integrations in \eqref{pi} and \eqref{pitilde}, we obtain
\begin{equation}
\omega^{(1)}(x)=\omega^{(2)}(x)=\frac{\mathrm{e}^{-\frac{1}{4}x^2}}{2\sqrt{\pi}}\ .\label{defpigauss}
\end{equation}

\begin{figure}
	\centering
	\includegraphics[scale=0.7]{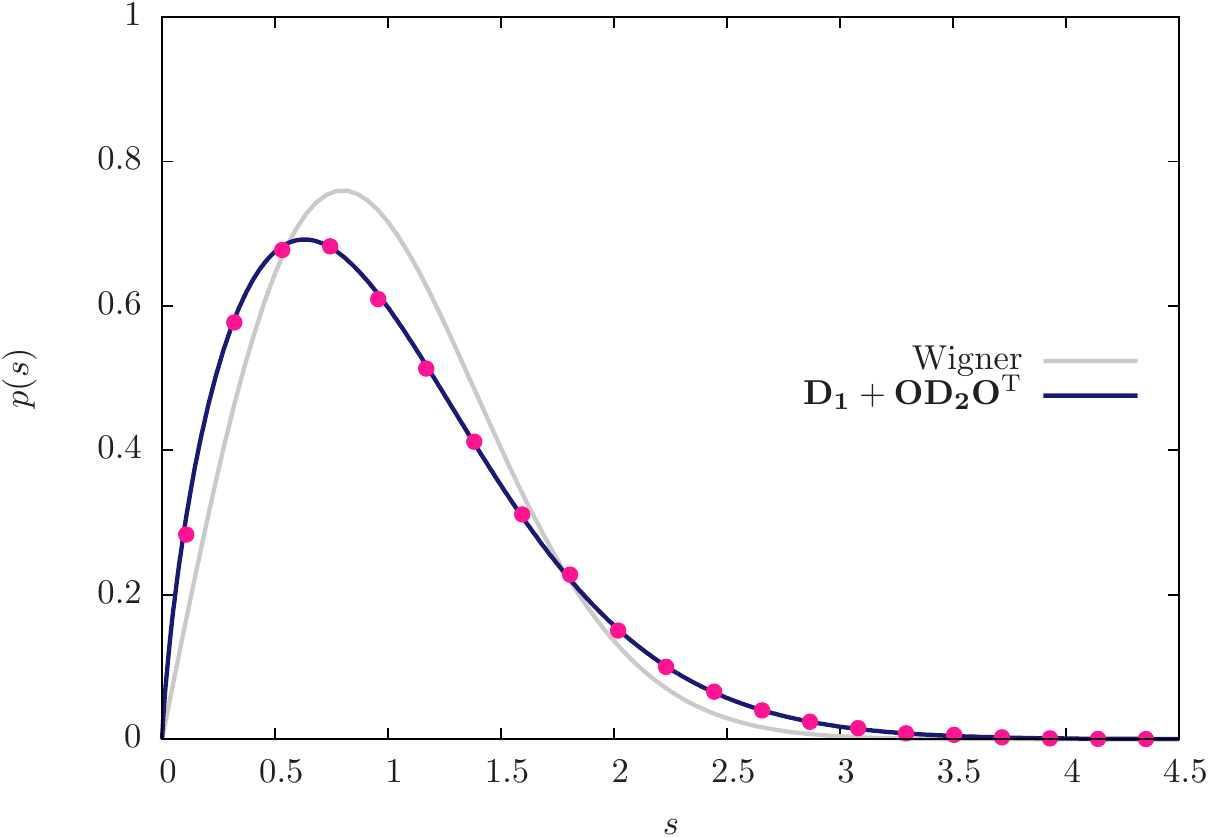}
	\caption{Spacing distribution for $2 \times 2$ random matrices formed as $\mathbf{D}_1 + \mathbf{O} \mathbf{D}_2 \mathbf{O}^\mathrm{T}$ (circles), where $\mathbf{D}_{1,2}$ are diagonal matrices whose diagonal entries are i.i.d. variables drawn from a standard Gaussian distribution, whereas $\mathbf{O}$ is a random Haar orthogonal matrix. The solid blue line refers to the analytical prediction \eqref{orth_spacing} for this quantity specialized to the Gaussian case (see equation \eqref{gaussfinalspacing}). The grey solid line shows Wigner's surmise \eqref{WS} for $\beta = 1$. Both the formula in equation \eqref{orth_spacing}, and the corresponding numerical results have been rescaled to have unit mean in order to be properly compared to Wigner's surmise. This can be easily achieved by mapping $p(s) \rightarrow \langle s \rangle p (\langle s \rangle s)$.}
	\label{spacing_2x2_orthogonal}
\end{figure}

Inserting this result into \eqref{fq} and invoking \eqref{anyj} the problem is reduced to the calculation of the following integral
\begin{align}
\nonumber \mathcal{P}(q) &=\frac{1}{2\pi}\times 4\times \frac{1}{(2\sqrt{\pi})^2}\int_{\sqrt{q}}^\infty \de x\int_{-\sqrt{q}}^{\sqrt{q}}\de y \frac{\mathrm{e}^{-\frac{1}{4}\left(\frac{x-y}{2}\right)^2-\frac{1}{4}\left(\frac{x+y}{2}\right)^2}}{\sqrt{(x^2-q)(q-y^2)}}\\
&=\frac{1}{\pi^2}\int_{\sqrt{q}}^\infty \de x \frac{\mathrm{e}^{-\frac{1}{8}x^2}}{\sqrt{x^2-q}}\int_0^{\sqrt{q}}\de y\frac{\mathrm{e}^{-\frac{1}{8}y^2}}{\sqrt{q-y^2}}\ ,
\end{align}
where in the last line we used the parity of the integrand. Performing the integrations we obtain
\begin{equation}
\mathcal{P}(q) =\frac{1}{4\pi}\mathrm{e}^{-q/8}I_0\left(\frac{q}{16}\right)K_0\left(\frac{q}{16}\right)\ ,
\end{equation}
where $I_0(x)$ and $K_0(x)$ are modified Bessel functions. Using \eqref{qtos} we get for the spacing distribution
\begin{equation}
p(s)=\frac{s}{2\pi}\mathrm{e}^{-s^2/8}I_0\left(\frac{s^2}{16}\right)K_0\left(\frac{s^2}{16}\right)\ ,\label{gaussfinalspacing}
\end{equation}
which is correctly normalized, $\int_0^\infty \de s\ p(s)=1$. In Fig. \ref{spacing_2x2_orthogonal} we show excellent agreement between equation\eqref{gaussfinalspacing} and the spacing distribution of numerically generated random matrices. The behavior as $s\to 0^+$ is indeed as given in \eqref{genasy}
\begin{equation}
p(s)\sim -\frac{1}{\pi}s\ln s\ ,
\end{equation}
since $\omega^{(1)}(0)=1/2\sqrt{\pi}$ from \eqref{defpigauss}.

\subsubsection{Unitary case}\label{unitary2}
We can rewrite \eqref{unitpq} as
\begin{align}
\nonumber\mathcal{P}(q) =2  \int_{-\infty}^\infty \de s_1 \de s_2\int_0^{\pi/2}\!\!\!\de t\sin(t)\cos(t)  \delta\left(q-\left[s_1^2+s_2^2+2 s_1 s_2\cos(2t)\right]\right)\omega^{(1)}(s_1)\omega^{(2)}(s_2)\ .\\
\end{align}
Repeating the same steps as in the orthogonal case, if $0<\omega^{(1)}(0)<\infty$, we find now that for $M=N=2$ the level repulsion at zero is universally given by the following non-Wigner behavior 
\begin{equation}
p(s)\sim \pi^2 [\omega^{(1)}(0)]^2 s\ ,\qquad s\to 0^+\ ,\label{genasyunitary}
\end{equation}
i.e. we observe a \emph{linear} repulsion instead of the quadratic behavior one normally expects for unitarily invariant ensembles.

In order to prove \eqref{genasyunitary}, we start from
\begin{equation}
\label{unit_spacing}
p(s)= 2 s \int_s^\infty \de x\int_{-s}^s \de y \frac{2\omega^{(1)}\left(\frac{x-y}{2}\right)\omega^{(1)}\left(\frac{x+y}{2}\right)}{x^2-y^2}\ .
\end{equation}
For small $s$, the integral in $y$ can be estimated as
\begin{align}
\nonumber\int_{-s}^s \de y \frac{\omega^{(1)}\left(\frac{x-y}{2}\right)\omega^{(1)}\left(\frac{x+y}{2}\right)}{x^2-y^2}\to \left[\omega^{(1)}\left(\frac{x}{2}\right)\right]^2 \int_{-s}^s \frac{\de y}{x^2-y^2}=2\left[\omega^{(1)}\left(\frac{x}{2}\right)\right]^2\frac{\mathrm{arctanh}(s/x)}{x}\ .\\
\end{align}
Therefore
\begin{equation}
p(s)\sim 8s \int_s^\infty \de x \left[\omega^{(1)}\left(\frac{x}{2}\right)\right]^2 \frac{\mathrm{arctanh}(s/x)}{x}\ .\label{intx}
\end{equation}
For $s$ close to zero we can estimate the integral contribution as
\begin{align}
\nonumber\int_s^\infty \de x \left[\omega^{(1)}\left(\frac{x}{2}\right)\right]^2 \frac{\mathrm{arctanh}(s/x)}{x}\to \left[\omega^{(1)}\left(0\right)\right]^2\int_s^1\de x\frac{\mathrm{arctanh}(s/x)}{x}\stackrel{s\to 0^+}{\longrightarrow}\frac{\pi^2}{8} \left[\omega^{(1)}\left(0\right)\right]^2\ ,\\
\end{align}
directly yielding \eqref{genasyunitary}. At the end of next subsection, we shall recover this behavior from an explicit calculation of the spacing distribution for Gaussian distributed eigenvalues, starting this time from the jpd of eigenvalues \eqref{jpdeigfinaltext}.
\begin{figure}
	\centering
	\includegraphics[scale=0.7]{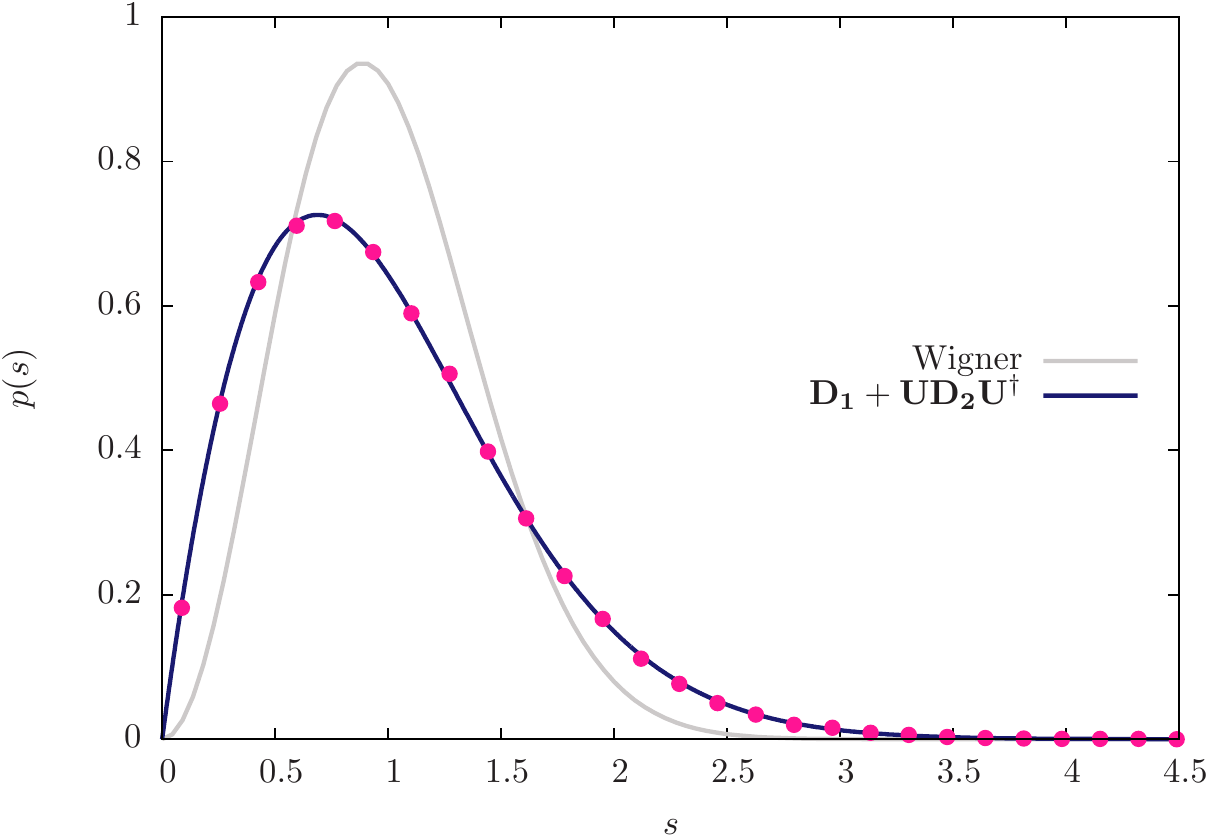}
	\caption{Spacing distribution for $2 \times 2$ random matrices formed as $\mathbf{D}_1 + \mathbf{U} \mathbf{D}_2 \mathbf{U}^\dagger$ (circles), where $\mathbf{D}_{1,2}$ are diagonal matrices whose diagonal entries are i.i.d. variables drawn from a standard Gaussian distribution, whereas $\mathbf{U}$ is a random Haar unitary matrix. The solid blue line refers to the analytical prediction \eqref{unit_spacing} for this quantity specialized to the Gaussian case (see equation \eqref{precisespacingunitary}). The grey solid line shows Wigner's surmise \eqref{WS} for $\beta = 2$. Both the formula in equation \eqref{unit_spacing}, and the corresponding numerical results have been rescaled to have unit mean in order to be properly compared to Wigner's surmise. This can be easily achieved by mapping $p(s) \rightarrow \langle s \rangle p (\langle s \rangle s)$.}
	\label{spacing_2x2_unitary}
\end{figure}

\subsection{From the jpd of eigenvalues}
\label{appA1}
Let us consider again the case $N=M=2$, this time for a standard Gaussian distribution 
$p^{(i)}(x)=\mathrm{e}^{-x^2/2}/\sqrt{2\pi},\quad\forall i$. We should then recover from \eqref{jpdeigfinalallequalsec} the results in previous subsection, specialized to the Gaussian law.

We get for $N=2$ (ignoring prefactors) 
\begin{equation}
\mathcal{P}(\nu_1,\nu_2)  \propto (\nu_2-\nu_1)\int_{-\infty}^\infty\de t_1\de t_2\de\lambda_1 \de\lambda_2 \de\ell_1 \de\ell_2
\exp\left(-\frac{\lambda_1^2}{2}-\frac{\lambda_2^2}{2}-\frac{\ell_1^2}{2}-\frac{\ell_1^2}{2}\right)
\frac{\psi(t_1,t_2,\lambda_1,\lambda_2,\ell_1,\ell_2,\nu_1,\nu_2)}{(t_2-t_1)(\lambda_2-\lambda_1)(\ell_2-\ell_1)}\ ,
\end{equation}
where
\begin{align}
\nonumber
\psi(t_1,t_2,\lambda_1,\lambda_2,\ell_1,\ell_2,\nu_1,\nu_2) & =\left[\mathrm{e}^{-\mathrm{i}(\nu_1 t_1+\nu_2 t_2)}-\mathrm{e}^{-\mathrm{i}(\nu_1 t_2+\nu_2 t_1)}\right]\left[\mathrm{e}^{\mathrm{i}(\lambda_1 t_1+\lambda_2 t_2)}-\mathrm{e}^{\mathrm{i}(\lambda_1 t_2+\lambda_2 t_1)}\right]\times \\ &\times \left[\mathrm{e}^{\mathrm{i}(\ell_1 t_1+\ell_2 t_2)}-\mathrm{e}^{\mathrm{i}(\ell_1 t_2+\ell_2 t_1)}\right]\ .
\end{align}

After lengthy algebra, we obtain
\be
\mathcal{P}(\nu_1,\nu_2)\propto (\nu_2-\nu_1)I(\nu_1,\nu_2)\ ,
\ee
where
\begin{equation}
 I(\nu_1,\nu_2)  = \int_{-\infty}^\infty\frac{\de t_1\de t_2 \left(\mathrm{e}^{-\mathrm{i}(\nu_1 t_1+\nu_2 t_2)}-\mathrm{e}^{-\mathrm{i}(\nu_1 t_2+\nu_2 t_1)}\right)}{t_2-t_1}\left[\int_{-\infty}^\infty\frac{\de x\de y}{y-x}\exp\left(-\frac{1}{2}(x^2+y^2)\right)\left(\mathrm{e}^{\mathrm{i}(x t_1+y t_2)}-\mathrm{e}^{\mathrm{i}(x t_2+y t_1)}\right)\right]^2\ .\label{defI}
\end{equation}
The integral in square brackets can be performed making the change of variables $\tau=(x+y)/2$ and $\zeta=(x-y)/2$ yielding
\begin{equation}
\int_{-\infty}^\infty\frac{\de x\de y}{y-x}\exp\left(-\frac{1}{2}(x^2+y^2)\right)\left(\mathrm{e}^{\mathrm{i}(x t_1+y t_2)}-\mathrm{e}^{\mathrm{i}(x t_2+y t_1)}\right)=-2\mathrm{i}\pi^{3/2}\mathrm{e}^{-(1/4)(t_1+t_2)^2}\mathrm{erf}\left(\frac{t_1-t_2}{2}\right)\ ,\label{evaluation}
\end{equation}
where $\mathrm{erf}(z)=(2/\sqrt{\pi})\int_0^z\ \mathrm{e}^{-t^2}\de t$ is the error function. It is now convenient to introduce the distribution $p(s)$ of the spacing $s=\nu_2-\nu_1$ between the two eigenvalues
\be
p(s)=2\int_{-\infty}^\infty\de\nu_1\de\nu_2 \mathcal{P}(\nu_1,\nu_2)\delta(s-(\nu_2-\nu_1))=2\int_{-\infty}^\infty\de\nu \mathcal{P}(\nu,\nu+s)\ .
\ee
Performing the remaining integrals in \eqref{defI}, we are left with
\be
p(s)\propto s J_2(s)\ ,\label{ps2}
\ee
where

\be
J_2(s)=\int_{-\infty}^{\infty} \de t \ \frac{\sin(s t)}{t} \left ( \mathrm{erf}(t) \right )^2=2\int_0^\infty \de t \ \frac{\sin(s t)}{t} \left ( \mathrm{erf}(t) \right )^2\ .\label{J2s}
\ee
In \eqref{J2s}, the term $(\mathrm{erf}(t))^2$ descends from the double integral evaluation in \eqref{evaluation} which needs to be squared (see \eqref{defI}). The term $\sin(s t)$ instead comes from the phase difference in \eqref{defI}.

The integral $J_2(s)$ can be written in the alternative and more convenient form
\be
J_2(s)=4\int_0^{\pi/4}\de\theta\ \mathrm{erfc}\left(\frac{s}{2}\cos\theta\right)\ ,\label{J2sbis}
\ee
where $\mathrm{erfc}(z)=1-\mathrm{erf}(z)$ is the complementary error function. To go from \eqref{J2s} to \eqref{J2sbis}, one first writes $(\mathrm{erf}(t))^2=(4/\pi)\int_{(0,t)^2}\de z_1\de z_2 \exp(-z_1^2-z_2^2)$, then performs the change of variables $z_{1,2}=t\zeta_{1,2}$. The integral in $t$ can be performed first, and the remaining integrals in $\zeta_{1,2}$ from $0$ to $1$ can be solved in polar coordinates, leaving eventually the angular integral in \eqref{J2sbis} which cannot be evaluated in closed form.

This representation allows to fix the normalization of $p(s)$ using
\begin{align}
&\int_0^\infty\de s\ s\ \mathrm{erfc}\left(\frac{s}{2}\cos\theta\right) =\frac{1}{(\cos\theta)^2}\ ,\\
&\int_0^{\pi/4}\de\theta \frac{1}{(\cos\theta)^2} =1\ .
\end{align}
Eventually we obtain precisely
\be
p(s)=s \int_0^{\pi/4}\de\theta\ \mathrm{erfc}\left(\frac{s}{2}\cos\theta\right)\ ,\label{precisespacingunitary}
\ee
(normalized to $1$) whose behavior for $s\to 0^+$ is $p(s)\sim (\pi/4)s+\cdots$, in agreement with the general result \eqref{genasyunitary}. In Fig. \ref{spacing_2x2_unitary} we show a perfect agreement between numerically generated matrices and the spacing distribution \eqref{precisespacingunitary}.

Using the previous results, we can now tackle another case, namely the complex hermitian (unitary) case with $N=2$ and $M>2$. Even with the smallest possible size $(N=2)$, this case cannot be efficiently dealt with as in Section \ref{unitary2}. However, we can still exploit the exact jpd \eqref{jpdeigfinalallequal} and the integral \eqref{defI} (with the exponent of the square bracket replaced by $M$ and $t_2-t_1$ replaced by $(t_2-t_1)^{M-1}$) to get for the spacing distribution
\be
p(s)=K_M s J_M(s)\ ,\label{psM}
\ee
where
\be
J_M(s)=\int_0^\infty \de t \ \frac{\sin(s t)}{t^{M-1}} \left [ \mathrm{erf}(t) \right ]^M\ .\label{JMSapp}
\ee
Note that for $M=2$, \eqref{psM} reproduces \eqref{ps2} as it should. For $M>2$, the algebraic manipulations yielding from \eqref{J2s} to \eqref{J2sbis} do not seem to work, therefore the normalization constant $K_M$ must be fixed case by case. However, it is quite easy to find out that for $M\geq 4$ the behavior for small spacings is \emph{quadratic}
\be
\label{psM_asy}
p(s)\sim K_M \omega_M s^2,\qquad s\to 0^+\ ,
\ee
with
\be
\omega_M=\int_0^\infty \de t\frac{[\mathrm{erf}(t)]^M}{t^{M-2}}\ .\label{omegaM}
\ee
In some sense, for $M\geq 4$ we recover a ``Wigner-like" behavior, even though the details of the spacing distribution are clearly different from the Wigner's surmise \eqref{WS} for $\beta=2$. In Sec. \ref{appvectorsU}, we have seen that for $M\gg 1$ the spacing distribution (for \emph{any} distribution of eigenvalues with finite variance $\sigma^2$) attains a scaling form $p(s)\to\frac{1}{\sigma\sqrt{M}}\Phi_{\mathbf{U}}\left(\frac{s}{\sigma\sqrt{M}}\right)$, where the scaling function $\Phi_{\mathbf{U}}(x)$ (see \eqref{scaling2}) is different from \eqref{WS} for $\beta=2$.

\begin{figure}
	\centering
	\includegraphics[scale=0.7]{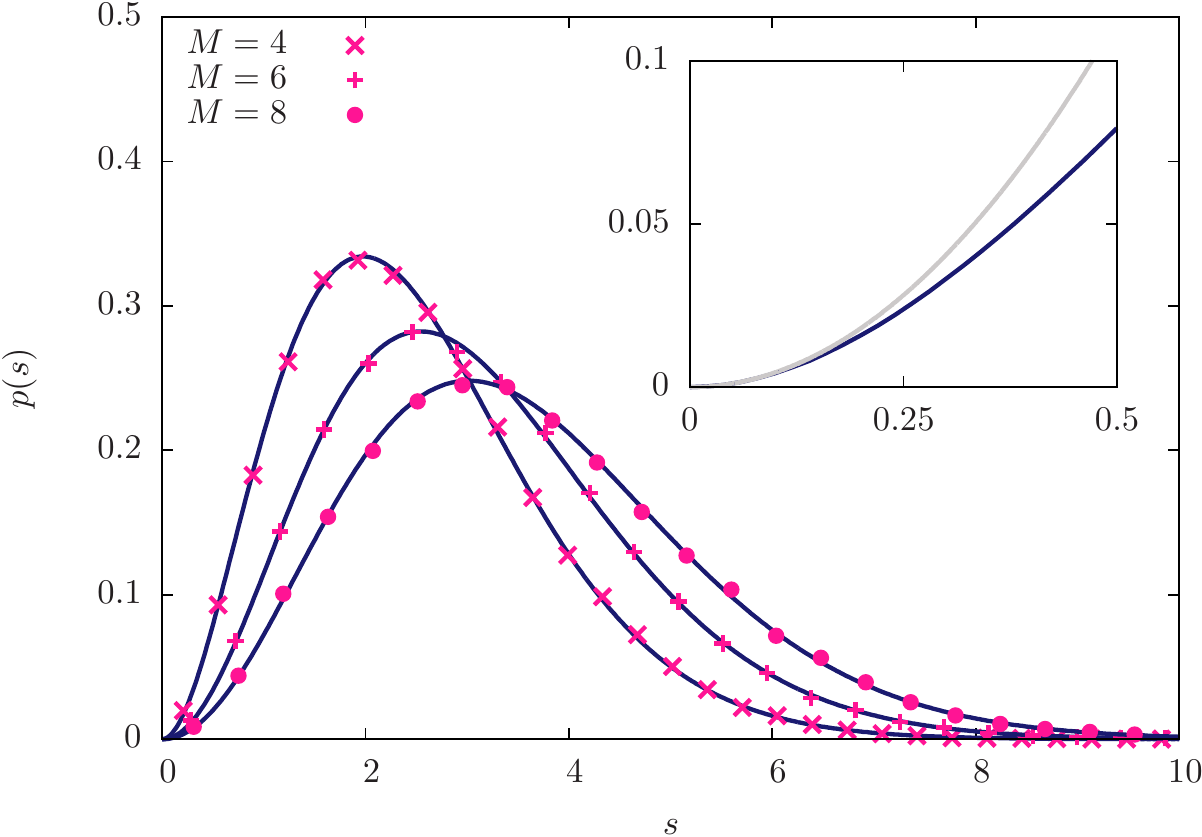}
	\caption{Comparison between equation \eqref{psM} and the distribution of eigenvalue spacings obtained numerically from the addition of $M = 4, 6, 8$ unitary $2 \times 2$ random matrices (see equation \eqref{defS}). The inset shows the comparison between the exact result of equation \eqref{psM} (dark blue line) and its asymptotic behavior for small $s$ \eqref{psM_asy} for the case $M = 4$.}
	\label{spacing_2x2_M}
\end{figure}

In Fig. \ref{spacing_2x2_M}, we include plots of the spacing distributions for different $M$ (this time, without adjusting the average $\langle s\rangle$ to $1$), together with numerical simulations.

The case $M=3$ is instead different (note that for $M=3$, the constant $\omega_3$ in \eqref{omegaM} would be divergent). The behavior for small spacing is quite exotic,
\be
\label{ps3_asy}
p(s)\sim -K_3 s^2\ln s,\qquad s\to 0^+\ ,
\ee
and is determined by the small $s$ behavior of the integral $J_3(s)$ from \eqref{JMSapp}, $J_3(s)\sim -s\ln s$ for $s\to 0^+$.

\vspace{0.05\textheight}

\end{document}